\documentclass[12pt]{article}
\usepackage{amssymb}
\usepackage{amsmath,bm}
\usepackage{graphics}
\usepackage{graphicx,epsfig}
\usepackage{subfigure}
\usepackage{placeins}
\usepackage{indentfirst}
\usepackage{setspace}
\usepackage{enumerate}
\usepackage{enumitem}
\usepackage{natbib}
\usepackage{caption}
\usepackage{varioref}
\usepackage{booktabs,tabularx}
\usepackage{color,float}
\usepackage{scalefnt}

\usepackage{epstopdf}
\usepackage{hyperref,amsfonts,dsfont}
\usepackage[title]{appendix}
\usepackage{amsthm}
\makeatletter \oddsidemargin  -.1in \evensidemargin -.1in
	\numberwithin{equation}{section}
	\renewcommand{\theequation}{\thesection.\arabic{equation}}
 
\begin{document}
	\newcommand{\bea}{\begin{eqnarray}}
		\newcommand{\eea}{\end{eqnarray}}
	\newcommand{\nn}{\nonumber}
	\newcommand{\bee}{\begin{eqnarray*}}
		\newcommand{\eee}{\end{eqnarray*}}
	\newcommand{\lb}{\label}
	\newcommand{\nii}{\noindent}
	\newcommand{\ii}{\indent}
	\numberwithin{equation}{section}
	\renewcommand{\theequation}{\thesection.\arabic{equation}}
     \onehalfspacing
	\title{\bf A New Generalized Birnbaum-Saunders Regression Model: Inference, Diagnostics, and Applications}
	\author{ Matheus B. Milhomem$^1$,~  Terezinha K. A. {Ribeiro}$^{1\,\!}$\thanks {Email address: terezinha.ribeiro@unb.br},\\~  Michelli {Barros}$^{2\,\!}$~~and~~Eriton B. {Santos}$^1$
		\\{\it \small $^1$Department of Statistics, University of Brasilia,  70910-900, Brazil}\\[-0,2cm]
        {\it \small  $^2$Department of Statistics, Federal University of Campina Grande,  58109-970, Brazil}
        }
        
\date{}
\maketitle
\begin{center}
	\textbf{Abstract}
\end{center}

The Birnbaum–Saunders (BS) distribution has become a widely used model for positive and asymmetric continuous data. Several extensions of the BS distribution have been proposed, including regression formulations that relate its parameters to covariates. In this paper, we introduce a new regression model for the Generalized Birnbaum–Saunders (GBS) distribution, an extension that has received limited attention in the literature despite offering a stronger physical interpretation. The proposed framework follows the structure of Generalized Additive Models for Location, Scale, and Shape (GAMLSS), allowing covariate effects to be incorporated into all three parameters of the distribution, thereby providing greater flexibility for modeling data. In particular, one of the model parameters has a direct interpretation as the median of the response variable, facilitating the interpretation of covariate effects on the conditional median. Parameter estimation is performed via maximum likelihood, and hypothesis tests are proposed based on the Wald test. The proposed regression model is implemented in \texttt{R} through the \texttt{gamlss} package, allowing access to a broad range of tools for model fitting, diagnostics, and assessment.  Monte Carlo simulation studies are conducted to investigate the finite-sample performance of the maximum likelihood estimators. The results show that the estimators exhibit desirable properties, with bias decreasing and efficiency increasing as sample size grows. Additionally, the behavior of the Wald test is investigated, showing good performance for moderate to large sample sizes. Finally, two applications to real data illustrate the practical usefulness of the proposed GBS regression model, demonstrating that it is a flexible alternative for modeling positive and asymmetric data compared to the BS model.

\paragraph{Keywords.} GBS distribution, Regression models, GAMLSS, Maximum likelihood estimation, Wald test, Quantile residuals.


\section{Introduction}
\noindent

The Birnbaum–Saunders (BS) probability distribution has been extensively studied due to its appealing mathematical properties and physical motivation based on Miner’s rule. Widely used in engineering, this rule treats fatigue as a deterministic phenomenon and relies on concepts such as accumulated damage and internal energy to describe its underlying mechanism, as discussed by \cite{birnbaum1968probabilistic}. Fatigue processes can be understood as the progressive accumulation of damage in a material or component resulting from repeated use or exposure to stress and strain, with failure occurring when the accumulated damage reaches a critical resistance threshold. The probabilistic interpretation of Miner’s rule provided by \cite{birnbaum1969new} led to the formal introduction of the BS distribution. Although originally developed for material fatigue, the BS distribution has since been applied in several fields, including business, environmental sciences, and health. For example, \cite{leiva2014capability} used the BS distribution to model protein requirements in restricted diets for hospitalized patients, \cite{leiva2016extreme} developed an extreme-value BS regression model for maximum daily ambient ozone concentration, \cite{leao2018survival} proposed a BS-based frailty model for cancer survival data, and \cite{sanchez2021birnbaum} developed a quantile regression framework for household income data.

Several extensions of the BS distribution have been proposed to increase its flexibility and physical interpretability. \cite{desmond1985stochastic} introduced a biologically motivated extension by relaxing the assumption of independence among stress cycles. \cite{owen1999accelerated} proposed a three-parameter generalization by incorporating a parameter related to the size of the material or component, arguing that size affects resistance and, consequently, the number of stress cycles that can be sustained. Other extensions include the elliptical BS distribution of \cite{diaz2002new}, a noncentral BS distribution proposed by \cite{guiraud2009non}, and a three-parameter class based on the scheme of \cite{marshall1997new} introduced by \cite{lemonte2013new}.

This paper focuses on the three-parameter BS generalization introduced by \cite{owen2006new}, which incorporates an additional parameter to account for dependence among the stress cycles experienced by a material or component. By relaxing the independence assumption, this formulation provides a more realistic physical interpretation of the fatigue process while also increasing  the flexibility of the BS distribution.

Regression models based on the BS distribution have also received considerable attention. \cite{rieck1991log} proposed a log-linear regression model based on the relationship between the BS and sinh-normal distributions, which was later generalized to nonlinear regression models by \cite{lemonte2009birnbaum}. \cite{leiva2016extreme} developed regression models based on the extreme-value BS distribution, including maximum likelihood estimation and residual analysis. More recently, \cite{bs_quantile_regression} and \cite{silva2022quantile} proposed quantile regression models based on the BS distribution and its generalizations. 

Alternative parameterizations of the BS distribution have also been investigated. \cite{santos2012new} studied eleven parameterizations, including nine newly proposed forms, and showed that one of them directly models the mean of a BS random variable on its original scale. Based on this parameterization, \cite{leiva2014birnbaum} developed a regression framework analogous to generalized linear models (GLMs) \citep{nelder1972generalized}. Subsequently, \cite{santos2016reparameterized} extended this framework by incorporating regression structures for both the mean and precision parameters, in a formulation analogous to double GLMs.

Despite these developments, to the best of our knowledge, no regression model has been proposed that directly uses the three-parameter BS distribution introduced by \cite{owen2006new} as the conditional distribution of the response variable. This gap is particularly relevant because the additional parameter of this distribution allows the dependence among stress cycles to be incorporated into the modeling framework, providing greater flexibility than regression models based on the classical BS distribution. Thus, introducing a regression model based on the GBS distribution not only extends the existing BS regression framework but also provides a useful tool for applications in which the assumption of independent stress cycles may be unrealistic. 

The main objective of this paper is to develop a new class of regression models based on the three-parameter BS distribution of \cite{owen2006new}, incorporating a structure analogous to the Generalized Additive Models for Location, Scale, and Shape (GAMLSS) framework introduced by \cite{rigby2005generalized}. Parameter inference is performed using maximum likelihood, and the finite-sample behavior of the estimators and Wald tests is investigated through simulation studies. The applicability of the proposed models is illustrated through two applications to real datasets. To facilitate reproducibility and practical applications, all code is publicly available in \texttt{R} through the GitHub repository \url{https://github.com/braga0m/GBS_Regression_Models}. The simulations and applications were implemented using the architecture of the \texttt{gamlss} library proposed by \cite{rigby2005generalized}.

The remainder of this paper is organized as follows. Section \ref{sec2} presents the GBS distribution and some of its main properties. Section \ref{sec3} introduces the proposed regression model. Section \ref{sec4} discusses inference, diagnostic procedures, and model selection. Section \ref{sec5} presents simulation studies evaluating the performance of the maximum likelihood estimators and Wald tests. Section \ref{sec6} illustrates the applicability of the proposed models through two real data applications. Finally, concluding remarks are presented in Section \ref{sec7}.

\section{The generalized Birnbaum Saunders distribution}\label{sec2}

 In the original derivation of the BS distribution, failure occurs when the cumulative crack growth reaches a critical threshold, and the increments produced by successive stress cycles are assumed to be independent and identically distributed, so that the classical central limit theorem applies. \cite{owen2006new} relaxes the independence assumption by modeling the sequence of crack extensions as a dependent process with memory. The resulting cumulative crack growth is characterized through an appropriate asymptotic limit, leading to a generalized distribution with an additional parameter $0<\nu<1$, which controls the dependence structure among successive stress cycles. 

The BS distribution is recovered as the special case when \(\nu = 0.5\). For $0.5 < \nu <1$, the crack-growth process exhibits long-memory dependence, meaning that the effects of past stress cycles persist over time and the cumulative process exhibits stronger dependence among successive cycles. In contrast, for $0<\nu<0.5$, the process exhibits short-memory dependence, indicating that the influence of past stress cycles decays more rapidly. Hence, the parameter $\nu$ determines the memory behavior of the underlying crack-growth process and distinguishes the GBS distribution from the original BS distribution.

 A random variable $Y$ is said to follow the distribution referred to as the generalized Birnbaum--Saunders (GBS) distribution throughout this paper, based on the formulation proposed by \citet{owen2006new}, if its cumulative distribution function (CDF) and probability density function (PDF) are given, respectively, by
\begin{equation}
    F(y; \mu, \sigma, \nu) = \Phi \left[\frac{1}{\sigma} \left(  \frac{y^{1 - \nu}}{\sqrt{\mu}} - \frac{\sqrt{\mu}}{y^{\nu}}\right)  \right], \quad y >0,
\label{cdf_gbs}
\end{equation}
\begin{equation}
   f(y;\mu, \sigma, \nu) = \frac{1}{\sqrt{2\pi}} \frac{1}{\sigma \sqrt{\mu}y^\nu} \left(1 -\nu + \frac{\nu\mu}{y}  \right) \text{exp} \left\{-\frac{1}{2\sigma^2}\frac{(y - \mu)^2}{\mu y^{2\nu}}  \right\}, \quad y>0,
   \label{gbs_pdf}
\end{equation}
where $\mu>0$, $\sigma>0$, $0<\nu<1$, and $ \Phi(\cdot)$ denote the CDF of the standard normal distribution. We denoted by $Y \sim \rm{GBS(\mu, \sigma, \nu)}$.


An important feature of the GBS distribution is that its parameter \(\mu\) coincides with the median. This follows directly from the CDF in \eqref{cdf_gbs}, since at $y=\mu$, $F(\mu; \mu, \sigma, \nu) = \Phi(0) = 0.5$. This property makes $\mu$ particularly useful as a measure of central tendency and provides a direct interpretation of the parameter independent of the values of $\sigma$ and \(\nu\). However, unlike the classical BS distribution, the parameter $\mu$ in the GBS model does not serve as a scale parameter. Instead, the three parameters $\mu$, $\sigma$, and $\nu$ jointly determine the shape of the distribution.

The widespread use of the BS distribution is largely due to its close relationship with the standard normal distribution. The GBS distribution shares this characteristic with the BS distribution. In particular, the quantiles of the GBS distribution can be directly related to the corresponding quantiles of the standard normal distribution. Let $z_p$ denote the $p$th quantile of the standard normal distribution, and \( y_p \) denote the $p$th quantile of the GBS distribution, for \( 0 < p < 1 \). These quantiles satisfy the nonlinear equation
\begin{equation}
\sigma\sqrt{\mu}z_p y_p^{\nu}+\mu-y_p=0,
\label{GBSandN}
\end{equation}
where $z_p=\Phi^{-1}(p)$ and $y_p=F^{-1}(p;\mu,\sigma,\nu)$, with $\Phi^{-1}(\cdot)$ denoting the quantile function of the standard normal distribution.

Equation \eqref{GBSandN} provides a convenient method for obtaining the quantiles of the GBS distribution from the corresponding standard normal quantiles. Since the equation involves only a one-dimensional nonlinear problem, it can be solved numerically for $y_p$ once the parameters \(\mu\), \(\sigma\), \(\nu\), and the quantile \( z_p \) are specified. This representation is particularly useful in computational procedures involving the GBS distribution. To this end, we use the function {\tt uniroot} from the {\tt stats} package in the software {\tt R}.  The same relationship can also be used to generate pseudo-random values from the GBS distribution by replacing $z_p$ with a realization $Z\sim N(0,1)$ and solving \eqref{GBSandN}.

\section{The GBS regression model}\label{sec3}

The GBS regression model is defined as follows. Let  $Y_1, Y_2, \ldots, Y_n$ be independent response variable observations with $Y_i \sim {\rm GBS}(\mu_i, \sigma_i, \nu_i)$, for $i=1, 2, \ldots, n$. If $\nu_i=0.5$, $\forall i=1,\ldots,n$, we have $Y_i\sim {\rm BS}(\mu_i, \sigma_i)$.  The model can be expressed in vector form as follows:
\begin{equation}
\boldsymbol{Y}\overset{\text{ind}}{\sim} {\rm GBS}(\boldsymbol{\mu},  \boldsymbol{\sigma}, \boldsymbol{\nu}),
\label{BSG_DIST}
\end{equation}
where
\begin{equation}
\begin{aligned}[b]
g_1(\boldsymbol{\mu}) = \boldsymbol{\eta}_1  = \boldsymbol{X}_1\boldsymbol{\beta} + s_{11}(\textbf{x}_{11}) + \dots + s_{1J_1}(\textbf{x}_{1J_1}),\\
g_2(\boldsymbol{\sigma})= \boldsymbol{\eta}_2  = \boldsymbol{X}_2\boldsymbol{\gamma}  + s_{21}(\textbf{x}_{21}) + \dots + s_{2J_2}(\textbf{x}_{2J_2}),\\
g_3(\boldsymbol{\nu})= \boldsymbol{\eta}_3= \boldsymbol{X}_3\boldsymbol{\lambda} + s_{31}(\textbf{x}_{31}) + \dots + s_{3J_3}(\textbf{x}_{3J_3}),
\label{gamlss_general}
\end{aligned}
\end{equation}
with $\boldsymbol{\mu},  \boldsymbol{\sigma}$ and $\boldsymbol{\nu}$ are $n$-dimensional vectors of unkown parameters, $\boldsymbol{\eta}_k$, $g_k(.)$ and $\boldsymbol{X}_k$ are, respectively, predictor, known monotonic link function,  and design matrix associated with the $k$th parameter, where $k = 1, 2, 3$.  The vectors $\boldsymbol{\beta} \in \mathbb{R}^{p_1}, \boldsymbol{\gamma} \in \mathbb{R}^{p_2}$ and $\boldsymbol{\lambda} \in \mathbb{R}^{p_3}$ are the linear coefficient parameter associated with $\boldsymbol{\mu},  \boldsymbol{\sigma}$ and $\boldsymbol{\nu}$, respectively. Additionally, $s_{kj}(\textbf{x}_{kj})$  represent smoothing functions for covariates $\textbf{x}_{kj}$, with $j=1, \ldots, J_k$. 

The parameters of the GBS distribution are linked to predictors through regression structures analogous to those proposed in the GAMLSS methodology of \citet{rigby2005generalized}. This motivated the implementation of the GBS distribution within the GAMLSS framework.

The GAMLSS methodology provides substantial modeling flexibility by allowing the inclusion of linear and nonlinear parametric effects as well as nonparametric smooth functions of covariates. Consequently, a variety of additive structures may be specified, including penalized B-splines, monotone P-splines, cycle P-splines, varying coefficient P-splines, cubic smoothing splines, loess curve fitting, fractional polynomials, random effects, ridge regression and nonlinear parametric fits.

There are several possible choices for each link function. For instance, an appropriate choice for link function for \( g_1(\cdot) \) and \( g_2(\cdot) \) is the log link function, since \(\mu > 0\) and \(\sigma > 0\). In contrast, the logit link is a  common choice for  \( g_3(\cdot) \) as \(0 < \nu < 1\). Other link functions that can be used include identity, square root, inverse, and probit, among others.

Setting \( s_{kj}(\mathbf{x}_{kj}) = \boldsymbol{0} \) for $j=1,2,3$ and $j=1, \ldots, J_k$, i.e., without smoothing functions, we obtain the full parametric GBS regression model; otherwise, we consider the semiparametric form.

Our main interest in this paper is to propose, study, and apply fully parametric GBS regression models to both real and simulated data. Since the BS model is a special case of the GBS model, our implementation also supports fitting BS regression models within the GAMLSS framework. In this work, the comparison between GBS and BS regression models is conducted through a real data application presented in Section \ref{sec5}. In both GBS and BS regression models, the median response $\mu$ is modeled. However, GBS models include three parameters that can each have assigned regression structures, whereas BS models can have at most two regression submodels. Modeling $\sigma$ and $\nu$ as functions of covariates enhances the flexibility of the model. However, the corresponding coefficients $\gamma_j$ and \(\lambda_j\) lack direct practical interpretation because \(\sigma\) and \(\nu\) are shape parameters rather than quantities with straightforward meanings. Therefore, when the goal is to improve model flexibility rather than to interpret every regression coefficient, we recommend incorporating regression structures for all parameters of the GBS model. In this context, the coefficients $\beta_j$, $j=1,\ldots,p_1$,  have a direct interpretation in terms of the conditional median of the response, whereas the coefficients associated with the other parameters contribute to modeling the dispersion and dependence structure.


%

\section{Inference, diagnostics and model selection}\label{sec4}

Let $Y_1, \ldots, Y_n$ independent random variables with $\rm{GBS}$ distribution  with $Y_i \overset{ \rm ind}{\sim} \rm{GBS}(\mu_i, \sigma_i, \nu_i)$ and consider $\boldsymbol{\theta} = (\boldsymbol{\beta}^{\top}, \boldsymbol{\gamma}^{\top}, \boldsymbol{\lambda}^{\top})^\top \in \mathbb{R}^{p}$ with $p=p_1 + p_2 + p_3$.

The log-likelihood function for the fully parametric GBS regression models is given by
\begin{equation}
\begin{aligned}[b]
\ell(\boldsymbol{\theta}) = \sum_{i=1}^{n} \ell_i(\boldsymbol{\theta}),
\label{bsg_log_vero}
\end{aligned}
\end{equation}
with  
\begin{eqnarray*}
    \ell_i(\boldsymbol{\theta}) = -\frac{1}{2} \log(2\pi) -\log(\sigma_i) -\frac{1}{2} \log(\mu_i) - \nu_i\log(y_i) + \log\left(1 - \nu_i + \frac{\nu_i\mu_i}{y_i}\right) - \frac{1}{2}\frac{(y_i - \mu_i)^2}{\sigma_i^2 \mu_i y_i^{2\nu_i}}.
\end{eqnarray*}
The maximum likelihood estimator (MLE) is obtained by maximizing the log-likelihood function in (\ref{bsg_log_vero}) with respect to $\boldsymbol{\theta}$. The MLE for $\boldsymbol{\theta}$ is denoted by $\widehat{\boldsymbol{\theta}}$.

The score function, obtained by differentiating the
log-likelihood function given with respect to $\boldsymbol{\theta}$, is given by ${\boldsymbol U}(\boldsymbol{\theta}) = ({\boldsymbol U}_{\boldsymbol{\beta}}^\top, {\boldsymbol U}_{\boldsymbol{\gamma}}^\top, {\boldsymbol U}_{\boldsymbol{\lambda}}^\top)$, with
$$ U_{\boldsymbol{\beta}} = \boldsymbol{X}_1^{\top}T_1\boldsymbol{a},$$
$$ U_{\boldsymbol{\gamma}} = \boldsymbol{X}_2^{\top}T_2\boldsymbol{b},$$

$$ U_{\boldsymbol{\lambda}} = \boldsymbol{X}_3^{\top}T_3\boldsymbol{c},$$
where  $T_1= {\rm diag}\{1/g_1'(\mu_1),\ldots,1/g_1'(\mu_n)\}$, $T_2 = {\rm diag} = \{1/g_2'(\sigma_1),\ldots,1/g_2'(\sigma_n) \}$, $T_3 = {\rm diag}\{1/g_3'(\nu_1), ...,1/g_3'(\nu_n) \}$ are  $n\times n$ diagonal matrices,  $ \boldsymbol{a} = (a_1,  \ldots, a_n)^\top$, $\boldsymbol{b} = (b_1,  \ldots, b_n)^\top $, $\boldsymbol{c} = (c_1,  \ldots, c_n)^\top $  are $n$-dimensional vectors with
\begin{eqnarray*}
a_i= -\frac{1}{2\mu_i} + \frac{\nu_i}{y_i(1 - \nu_i) + \nu_i\mu_i} + \frac{y_i^2 - \mu_i^2}{2\sigma^2_i y_i^{2\nu_i}\mu_i^2},
\end{eqnarray*}
\begin{eqnarray*}
b_i = -\frac{1}{\sigma_i} + \frac{(y_i - \mu_i)^2}{\mu_iy_i^{2\nu_i}\sigma_i^3},
\end{eqnarray*}
\begin{eqnarray*}
c_i = (1 - \log(y_i)) \left(\frac{(y_i - \mu_i)^2}{\sigma_i^2 \mu_i y_i^{2\nu_i}} + \frac{\mu_i - y_i}{y_i(1 - \nu_i + \nu_i\mu_i)} \right).
\end{eqnarray*}

The MLE of $\boldsymbol{\theta}$ can be obtained by solving the system ${\boldsymbol U}(\boldsymbol{\theta}) = \boldsymbol{0}$, where $\boldsymbol{0}$ represents the $p$ dimensional null vector. However, this system cannot be solved analytically, and iterative numerical methods must be employed to obtain an approximate solution. 

To obtain the maximum likelihood estimates, we use the {\tt gamlss} function in the package with the same name \citep{stasinopoulos2008generalized} contained in {\tt R} \citep{r2025}. The GAMLSS framework proposed by \cite{rigby2005generalized} considers maximum likelihood estimation procedure based on the Rigby and Stasinopoulos (RS) algorithm and a generalization of the Cole and Green (CG) algorithm \citep{cole1992smoothing}.

Regarding computational aspects, the {\tt gamlss} function requires initial values for the parameters. Therefore,  we set the initial values as $\mu_i^{(0)} = \tilde{\mu}$, $ \sigma_i^{(0)} = 1.0$, and $\nu_i^{(0)} = 0.5$, where $\tilde{\mu}$ is the sample median of the response variable.
The set of starting values for the regression coefficients corresponds only to the intercepts of the regression submodels, while the remaining coefficients associated with the covariates are initialized to zero. Thus, the initial values for the intercepts of the $\mu$, $\sigma$, $\nu$ submodels are given by $\beta_1^{(0)}= g_1(\tilde{\mu})$, $\gamma_1^{(0)}= g_2(1.0)$, $\lambda_1^{(0)}= g_3(0.5)$, respectively.
Note that the initial values assume the response variable is independent and identically distributed. Despite their simplicity, based on our experience, these starting values are highly effective. We also tested another set of starting values derived from other fitted BS regression models; however, this did not lead to improvements in the simulation or application results.
Regarding the iterative algorithms available in GAMLSS, our experience indicates that the RS algorithm generally performs better. However, the CG algorithm also performs well in the majority of cases. 

With the implementation of the GBS regression models proposed in this work within the {\tt gamlss} framework, the BS regression model can also be fitted as a special case by setting $\nu=0.5$. Specifically, the BS regression model can be fitted using the {\tt gamlss} function by setting \texttt{nu.fix = TRUE} and \texttt{i.nu = 0.5}.

To construct confidence intervals and test hypotheses about the regression coefficients, we rely on the asymptotic normality of the MLE.
Under the standard regularity conditions for maximum likelihood estimation, when the sample size is large,
$$\widehat{\boldsymbol{\theta}} \stackrel{ {\rm a} }{ \sim } {\rm N}_{p}(\boldsymbol{\theta}, J^{-1}(\boldsymbol{\theta}) ),$$ where $\stackrel{ {\rm a} }{ \sim }$ denotes an approximate distribution, ${\rm N}_{p}(\boldsymbol{\theta}, J^{-1}(\boldsymbol{\theta}) )$ denotes the $p$-dimensional normal distribution with  mean ${\boldsymbol{\theta}}$ and variance–covariance matrix $J^{-1}(\boldsymbol{\theta})$, and $p$ represents the number of parameters in the model. The matrix 
$J^{-1}(\boldsymbol{\theta})$ is also the inverse of the observed information matrix. In this work, the observed information matrix is obtained numerically.

Under GBS regression models, we may be interested in evaluating the significance of the regression coefficients; that is, we want to test the hypothesis \(H_0: \theta_j = 0\) versus \(H_1: \theta_j \neq 0\), where \(\theta_j\) is the \(j\)-th element of \(\boldsymbol{\theta}\). To this end, we will use the Wald test  \citep{wald1943}, defined as follows.

Consider $H_0$: $\theta_j = \theta_j^0$ versus $H_1$: $\theta_j \neq \theta_j^0$, where $\theta_j^0$  is a specified value for the unknown parameter \(\theta_j\), with \( j = 1,  \ldots, p \). Under $H_0$,  $$z_j = \frac{\widehat{\theta}_j - \theta_j^0}{{\rm se}(\widehat{\theta}_j)} \stackrel{ {\rm a} }{ \sim } {\rm N}(0,1),$$
where $\widehat{\theta}_j$ is the MLE of $\theta_j$, and ${\rm se}(\widehat{\theta}_j) = \sqrt{J_{jj}}$ is the asymptotic standard error of $\widehat{\theta}_j$, with $J_{jj}$  the \(j\)-th diagonal element of \(J^{-1}(\boldsymbol{\theta})\). 
The test statistic $z_j$ is the square root of the Wald statistic, which is widely used in general regression models. 

 In particular, consider the GBS model, regardless of the specifications adopted for the $\mu$ and $\sigma$ submodels, with the $\nu$ submodel containing only the intercept and using the logit link, such that
 $\nu_i = \exp(\gamma_1)/(1+\exp(\gamma_1))$. In this setting, testing $H_0:\gamma_1=0$ versus $H_1:\gamma_1\neq 0$ is equivalent to testing $H_0$: $\nu_i = 0.5$ versus $H_1$: $\nu_i \neq 0.5$. Thus, a Wald test can be employed to determine whether there is evidence that $\nu$ differs from $0.5$, and consequently, whether the additional shape parameter of the GBS model is necessary to adequately describe the data. Since $\nu=0.5$ corresponds to the BS model, rejection of $H_0$ provides evidence in favor of the GBS specification over the corresponding BS regression model. This test will be used to compare the BS and GBS regression models in the application section.

To assess the goodness of fit under GBS regression models, the residuals introduced by \cite{dunn1996randomized}  can be employed.
The quantile residual $r_i^q$, $ i =1, \ldots, n$, is given by 
\begin{eqnarray*}\label{residuals}
r_i^q =   \Phi^{-1}(F(y_i; \widehat{\mu}_i,\widehat{\sigma}_i, \widehat{\nu}_i)).
\end{eqnarray*}
 If the GBS regression model is well-fitted, $r_i^q$ is approximated distributed as a standard normal distribution.
 To assess whether the empirical distribution of quantile residuals approximates a standard normal distribution, a common practice within the GAMLSS framework is to use the worm plot, as proposed by \citep{buuren2001worm}. This plot is a de-trended Q-Q plot of the residuals. If the GBS regression model fits the data well, the points in the worm plot should lie close to the middle horizontal line, with 95\% of them lying between the upper and lower dotted curves, indicating no systematic departure \citep{stasinopoulos2017flexible}. 
Other diagnostic tools commonly used within the GAMLSS framework include Q--Q plots, histograms, and kernel density estimates of the normalized quantile residuals. We also consider the Q--Q boxplot proposed by \citet{rodu2022q}, which provides an additional graphical assessment of residual normality, as employed by \citet{deCarvalho2026, otiniano2026regression, dutta2026}. This diagnostic plot combines information from the tails of a Q--Q plot with a traditional boxplot by replacing the boxplot whiskers with the corresponding Q--Q plot tails and displaying them alongside confidence bands. These graphical diagnostics provide complementary information for evaluating the adequacy of the fitted GBS regression model.

For model selection,  the Akaike Information Criterion (AIC) \citep{Akaike1974}, defined as $ {\rm AIC} =-2\ell(\widehat{\boldsymbol{\theta}}) + 2p$, and the Bayesian Information Criterion (BIC) \citep{Schwarz1978}, defined as $ {\rm BIC} =-2\ell(\widehat{\boldsymbol{\theta}}) + p\log(n)$ can also be considered.
Given a set of candidate models for the data, the preferred GBS regression model is the one with the lowest AIC and BIC values.

\section{Monte Carlo simulation results}\label{sec5}

To evaluate the performance of the MLE under GBS regression models, Monte Carlo simulations considering various different scenarios were carried out. The sample sizes considered are $n = 40, 80, 160, 320$. For all the scenarios we employ
log link function in the $\mu$ and $\sigma$ regression submodels, and logit link for the $\nu$ regression submodel. The models include intercepts, i.e. $x_{i1} = 1$,  for all $i = 1, 2, \ldots, n$, and the other covariates are taken as random draws from a standard uniform,  and from Bernoulli distribution with probability of success $0.7$. The covariates were kept constant throughout the simulated samples. All the results are based on 5000 Monte Carlo replications and were carried out using the R software environment \citep{r2025}.

Ten different parameter configurations were considered, as described in Table \ref{tab:scenarios}.

\begin{table}[!ht]
\centering
\caption{Parameter configurations considered in the simulation study.}
\label{tab:scenarios}
\setlength{\tabcolsep}{4pt}
\renewcommand{\arraystretch}{1.15}
\resizebox{\textwidth}{!}{%
\begin{tabular}{c|p{3.5cm}|p{5.2cm}|c|c|c|c|c}
\hline
\textbf{Scenario} &
\textbf{Specification} &
\textbf{Parameter values} &
$\boldsymbol{\mu}$\textbf{ range} &
\textbf{median}($\boldsymbol{\mu)}$&
$\boldsymbol{\sigma}$ \textbf{ range} &
\textbf{median}($\boldsymbol{\sigma)}$ &
$\boldsymbol{\nu}$ \textbf{ range} \\
\hline

1 &
One covariate in $\mu$; $\sigma$ and $\nu$ constant &
$\beta_1=1.8,\ \beta_2=-2.0,\ \gamma_1=-1.0,$
$\lambda_1=-1.39$ &
$(0.84,5.96)$ & 2.48 &
0.37 & -- & 0.2 \\\hline

2 &
One covariate in $\mu$; $\sigma$ and $\nu$ constant &
$\beta_1=1.8,\ \beta_2=-2.0,\ \gamma_1=-1.0,$
$\lambda_1=0.0$ &
$(0.84,5.96)$ & 2.48 &
0.37 & -- & 0.5 \\\hline

3 &
One covariate in $\mu$; $\sigma$ and $\nu$ constant &
$\beta_1=1.8,\ \beta_2=-2.0,\ \gamma_1=-1.0,$
$\lambda_1=1.39$ &
$(0.84,5.96)$ & 2.48 &
0.37 & -- & 0.8 \\\hline

4 &
One covariate in $\mu$; $\sigma$ and $\nu$ constant &
$\beta_1=2.7,\ \beta_2=-3.1,\ \gamma_1=-1.3,$
$\lambda_1=-1.39$ &
$(0.69,14.54)$ & 3.73 &
0.27 & -- & 0.2 \\\hline

5 &
One covariate in $\mu$; $\sigma$ and $\nu$ constant &
$\beta_1=2.7,\ \beta_2=-3.1,\ \gamma_1=-1.3,$
$\lambda_1=0.0$ &
$(0.69,14.54)$ & 3.73 &
0.27 & -- & 0.5 \\\hline

6 &
One covariate in $\mu$; $\sigma$ and $\nu$ constant &
$\beta_1=2.7,\ \beta_2=-3.1,\ \gamma_1=-1.3,$
$\lambda_1=1.39$ &
$(0.69,14.54)$ & 3.73 &
0.27 & -- & 0.8 \\\hline

7 &
One covariate in $\mu$ and $\sigma$; $\nu$ constant &
$\beta_1=1.1,\ \beta_2=0.7,\ \gamma_1=-1.1,$
$\gamma_2=1.6,\ \lambda_1=-0.5$ &
$(3.02,6.01)$ & 4.11 &
$(0.34,1.62)$ & 0.68 & 0.38 \\\hline

8 &
One covariate in $\mu$ and $\sigma$; $\nu$ constant &
$\beta_1=0.15,\ \beta_2=0.1,\ \gamma_1=-0.9,$
$\gamma_2=1.3,\ \lambda_1=0.5$ &
$(1.16,1.28)$ & 1.22 &
$(0.41,1.47)$ & 0.73 & 0.62 \\\hline

9 &
Two covariates in $\mu$ and $\sigma$; one covariate in $\nu$ &
$\beta_1=1.1,\ \beta_2=-2.6,\ \beta_3=0.7,$
$\gamma_1=-1.2,\ \gamma_2=1.1,\ \gamma_3=0.5,$
$\lambda_1=-1.2,\ \lambda_2=1.3$ &
$(0.32,5.93)$ & 1.29 &
$(0.36,1.48)$ & 0.71 &
$(0.23,0.52)$ \\\hline

10 &
Two covariates in $\mu$ and $\sigma$; one covariate in $\nu$ &
$\beta_1=1.8,\ \beta_2=-2.0,\ \beta_3=1.0,$
$\gamma_1=-1.2,\ \gamma_2=1.3,\ \gamma_3=1.4,$
$\lambda_1=-1.2,\ \lambda_2=1.3$ &
$(1.09,16.20)$ & 4.59 &
$(0.37,4.43)$ & 1.62 &
$(0.23,0.52)$ \\

\hline
\end{tabular}%
}
\end{table}

The primary difference among Scenarios 1–6 lies in the values of $\nu$. Specifically, $\nu$ increases progressively from 0.2 to 0.8 across Scenarios 1–3 and 4–6. In all six scenarios, covariates are included only in the $\mu$ submodel, while the $\sigma$ and $\nu$ submodels contain only intercept terms. Scenarios 1–3 share the same regression coefficients and parameter settings for $\mu$ and \(\sigma\), whereas Scenarios 4–6 consider a different set of regression coefficients, resulting in a wider range and larger median values of $\mu$. 
In Scenarios 7 and 8, covariates are included in both the $\mu$ and $\sigma$ submodels, whereas the $\nu$ submodel contains only an intercept term. 
Finally, Scenarios 9 and 10 consider regression structures involving all three parameters of the BSG model. In these scenarios, two covariates are included in the $\mu$ and $\sigma$ submodels, while one covariate is included in the \(\nu\) submodel. Thus, Scenarios 9 and 10 represent the most complex regression structures considered in the simulation study.

Through 5000 Monte Carlo replicates, we obtain for each $\theta_t$, $t = 1,\ldots, p$, the  parameter estimate and calculate, from their simulated sample distribution, the bias ${\rm B}(\widehat{\theta}_t)$,  and the square root of the mean square error ${\rm RMSE}(\widehat{\theta}_t)$, defined, respectively, by
\begin{eqnarray*}
  {\rm B}(\widehat{\theta}_t) = \frac{1}{5000}\sum_{j=1}^{5000}\left(\widehat{\theta}_t^j - \theta_t\right) \ \ \  \mbox{and} \ \ \ 
   {\rm RMSE}(\widehat{\theta}_t) = \sqrt{\frac{1}{5000}\sum_{j=1}^{5000}\left(\widehat{\theta}_t^j - \theta_t\right)^2},  
\end{eqnarray*}
where $\widehat{\theta}_t^j$ is the MLE of the $t$th element of the parameter vector ${\boldsymbol\theta}$ in the $j$th Monte Carlo replicate, with  $j \in \{1,  \ldots, 5000\}$. Furthermore, we also obtain the mean of the asymptotic standard error (ASE) for each $\theta_t$ defined by
$$ {\rm ASE}(\widehat{\theta}_t) = \frac{1}{5000}\sum_{j=1}^{5000}{\rm se}(\widehat{\theta}_t^j),$$
where ${\rm se} (\widehat{\theta}_t^j)$ is the asymptotic standard error of the $t$th element of  ${\boldsymbol\theta}$ in the $j$th replicate.

Tables \ref{SimulationResults1} and \ref{SimulationResults2}  show the  bias, RMSE, and ASE of the MLEs for different sample sizes under Scenarios 1--10.
Overall, the results indicate that the MLEs perform well across the considered scenarios. In most cases, the bias, RMSE, and ASE decrease as the sample size $n$ increases, including in the more complex scenarios involving a larger number of parameters. 
The bias is generally small and tends toward zero as $n$ increases. This pattern is particularly evident for the regression coefficients in the $\mu$ submodel, whose biases are already relatively small for $n=40$ and become negligible for larger sample sizes. Somewhat larger finite-sample biases are observed for the coefficients in the $\sigma$ and $\nu$ submodels, particularly in Scenarios 9 and 10. For example, for $n=40$, the biases of $\gamma_1$ are $-0.2049$ and $-0.2429$ in Scenarios 9 and 10, respectively. However, these biases decrease substantially as the sample size increases, reaching $-0.0215$ and $-0.0281$ when $n=320$. Moreover, RMSE and ASE become increasingly close as the sample size grows. This indicates that the estimated asymptotic standard errors provide a good approximation to the empirical variability of the MLEs, particularly for moderate and large sample sizes. Together with the reduction in bias and estimation error, this behavior provides empirical evidence supporting the consistency and asymptotic normality of the MLEs under GBS regression models.

\begin{table}[!ht]
\centering
\caption{ Bias, RMSE and ASE under Scenarios 1--6.}
\label{SimulationResults1}
\renewcommand{\arraystretch}{0.9}

\scriptsize
\setlength{\tabcolsep}{3pt}
\begin{tabular}{lrrrrrrrrrrrr}
\toprule

& \multicolumn{12}{c}{\textbf{Scenario 1}}\\
\cmidrule(lr){2-13}

& \multicolumn{3}{c}{$n=40$}
& \multicolumn{3}{c}{$n=80$}
& \multicolumn{3}{c}{$n=160$}
& \multicolumn{3}{c}{$n=320$}\\

\cmidrule(lr){2-4}
\cmidrule(lr){5-7}
\cmidrule(lr){8-10}
\cmidrule(lr){11-13}

Par.
& Bias & RMSE & ASE
& Bias & RMSE & ASE
& Bias & RMSE & ASE
& Bias & RMSE & ASE\\

\cmidrule(lr){2-4}
\cmidrule(lr){5-7}
\cmidrule(lr){8-10}
\cmidrule(lr){11-13}

$\beta_1$
& $-0.0049$ & 0.0744 & 0.0721 
& $-0.0016$ & 0.0519 & 0.0508 
& $-0.0012$ & 0.0366 & 0.0361
& $-0.0007$ & 0.0258 & 0.0256\\

$\beta_2$
& 0.0011 & 0.1559 & 0.1482
& 0.0013 & 0.1082 & 0.1058 
& 0.0023 & 0.0771 & 0.0753
&$-0.0001$&0.0540 & 0.0536\\

$\gamma_1$
& $-0.0796$ & 0.1835 & 0.1601
& $-0.0360$ & 0.1155 & 0.1093
& $-0.0216$ & 0.0806 & 0.0762
& $-0.0108$ & 0.0544 & 0.0534 \\

$\lambda_1$
& 0.0339  & 0.8944 & 1.2574
&$-0.0585$& 0.6866 & 0.7810
&$-0.0087$& 0.4523 & 0.4649
&$-0.0023$& 0.3054 & 0.3092  \\
\midrule

& \multicolumn{12}{c}{\textbf{Scenario 2}}\\
\cmidrule(lr){2-13}

& \multicolumn{3}{c}{$n=40$}
& \multicolumn{3}{c}{$n=80$}
& \multicolumn{3}{c}{$n=160$}
& \multicolumn{3}{c}{$n=320$}\\

\cmidrule(lr){2-4}
\cmidrule(lr){5-7}
\cmidrule(lr){8-10}
\cmidrule(lr){11-13}

Par.
& Bias & RMSE & ASE
& Bias & RMSE & ASE
& Bias & RMSE & ASE
& Bias & RMSE & ASE\\


\cmidrule(lr){2-4}
\cmidrule(lr){5-7}
\cmidrule(lr){8-10}
\cmidrule(lr){11-13}

$\beta_1$
& $-0.0046$ & 0.1100 & 0.1051
& $-0.0008$ & 0.0772 & 0.0755
& $-0.0009$ & 0.0546 & 0.0539
& $-0.0008$ & 0.0385 & 0.0383 \\

$\beta_2$
& 0.0069 & 0.1938 & 0.1837
& 0.0027 & 0.1352 & 0.1320
& 0.0029 & 0.0948 & 0.0939
& 0.0006 & 0.0678 & 0.0668 \\

$\gamma_1$
& $-0.0579$ & 0.1788 & 0.1654
& $-0.0257$ & 0.1165 & 0.1139
& $-0.0167$ & 0.0817 & 0.0797
& $-0.0081$ & 0.0553 & 0.0559  \\

$\lambda_1$
& 0.0141  & 0.6457 & 0.6532
&$-0.0111$& 0.4017 & 0.4043
& 0.0084  & 0.2695 & 0.2732
& 0.0029  & 0.1842 & 0.1890  \\
\midrule

& \multicolumn{12}{c}{\textbf{Scenario 3}}\\
\cmidrule(lr){2-13}

& \multicolumn{3}{c}{$n=40$}
& \multicolumn{3}{c}{$n=80$}
& \multicolumn{3}{c}{$n=160$}
& \multicolumn{3}{c}{$n=320$}\\

\cmidrule(lr){2-4}
\cmidrule(lr){5-7}
\cmidrule(lr){8-10}
\cmidrule(lr){11-13}

Par.
& Bias & RMSE & ASE
& Bias & RMSE & ASE
& Bias & RMSE & ASE
& Bias & RMSE & ASE\\

\cmidrule(lr){2-4}
\cmidrule(lr){5-7}
\cmidrule(lr){8-10}
\cmidrule(lr){11-13}

$\beta_1$
& 0.0012 & 0.1579 & 0.1516
& 0.0032 & 0.1128 & 0.1091
& 0.0009 & 0.0784 & 0.0779
& 0.0002 & 0.0554 & 0.0553 \\

$\beta_2$
& 0.0089 & 0.2403 & 0.2305 
& 0.0027 & 0.1697 & 0.1649
& 0.0036 & 0.1178 & 0.1171
& 0.0008 & 0.0836 & 0.0830  \\

$\gamma_1$
& $-0.0279$ & 0.1631 & 0.1651
& $-0.0106$ & 0.1141 & 0.1139  
& $-0.0076$ & 0.0801 & 0.0795
& $-0.0036$ & 0.0554 & 0.0558 \\

$\lambda_1$
& $-0.0355$ & 0.6608 & 0.7503
& $-0.0432$ & 0.4485 & 0.4590 
& $-0.0125$ & 0.3031 & 0.3104
& $-0.0094$ & 0.2079 & 0.2133 \\
\midrule

& \multicolumn{12}{c}{\textbf{Scenario 4}}\\
\cmidrule(lr){2-13}

& \multicolumn{3}{c}{$n=40$}
& \multicolumn{3}{c}{$n=80$}
& \multicolumn{3}{c}{$n=160$}
& \multicolumn{3}{c}{$n=320$}\\

\cmidrule(lr){2-4}
\cmidrule(lr){5-7}
\cmidrule(lr){8-10}
\cmidrule(lr){11-13}

Par.
& Bias & RMSE & ASE
& Bias & RMSE & ASE
& Bias & RMSE & ASE
& Bias & RMSE & ASE\\


\cmidrule(lr){2-4}
\cmidrule(lr){5-7}
\cmidrule(lr){8-10}
\cmidrule(lr){11-13}

$\beta_1$
& $-0.0013$ & 0.0447 & 0.0433
& $-0.0002$ & 0.0313 & 0.0306
& $-0.0003$ & 0.0220 & 0.0218
& $-0.0001$ & 0.0156 & 0.0154 \\

$\beta_2$
& $-0.0002$ & 0.1063 & 0.1011
& 0.0002    & 0.0734 & 0.0723
& 0.0012    & 0.0524 & 0.0514
& $-0.0003$ & 0.0369 & 0.0365 \\

$\gamma_1$
& $-0.0735$ & 0.1955 & 0.1782
& $-0.0302$ & 0.1264 & 0.1228
& $-0.0197$ & 0.0895 & 0.0859
& $-0.0094$ & 0.0606 & 0.0604 \\

$\lambda_1$
& $-0.0473$ & 0.7873 & 0.9998
& $-0.0770$ & 0.5798 & 0.6047
& $-0.0159$ & 0.3708 & 0.3685
& $-0.0094$ & 0.2487 & 0.2497 \\
\midrule

& \multicolumn{12}{c}{\textbf{Scenario 5}}\\
\cmidrule(lr){2-13}

& \multicolumn{3}{c}{$n=40$}
& \multicolumn{3}{c}{$n=80$}
& \multicolumn{3}{c}{$n=160$}
& \multicolumn{3}{c}{$n=320$}\\

\cmidrule(lr){2-4}
\cmidrule(lr){5-7}
\cmidrule(lr){8-10}
\cmidrule(lr){11-13}

Par.
& Bias & RMSE & ASE
& Bias & RMSE & ASE
& Bias & RMSE & ASE
& Bias & RMSE & ASE\\

\cmidrule(lr){2-4}
\cmidrule(lr){5-7}
\cmidrule(lr){8-10}
\cmidrule(lr){11-13}

$\beta_1$
& $-0.0023$ & 0.0815 & 0.0781
& $-0.0004$ & 0.0577 & 0.0559
& $-0.0004$ & 0.0403 & 0.0399
& $-0.0004$ & 0.0287 & 0.0283 \\

$\beta_2$
& 0.0030 & 0.1449 & 0.1381
& 0.0011 & 0.1014 & 0.0988
& 0.0017 & 0.0710 & 0.0702
& 0.0002 & 0.0506 & 0.0499 \\

$\gamma_1$
& $-0.0571$ & 0.1949 & 0.1803
& $-0.0255$ & 0.1277 & 0.1243
& $-0.0170$ & 0.0887 & 0.0871
& $-0.0079$ & 0.0607 & 0.0612 \\

$\lambda_1$
& 0.0108    & 0.5105 & 0.4976
& $-0.0059$ & 0.3264 & 0.3235 
& 0.0084    & 0.2202 & 0.2217
& 0.0020    & 0.1520 & 0.1543 \\
\midrule

& \multicolumn{12}{c}{\textbf{Scenario 6}}\\
\cmidrule(lr){2-13}

& \multicolumn{3}{c}{$n=40$}
& \multicolumn{3}{c}{$n=80$}
& \multicolumn{3}{c}{$n=160$}
& \multicolumn{3}{c}{$n=320$}\\

\cmidrule(lr){2-4}
\cmidrule(lr){5-7}
\cmidrule(lr){8-10}
\cmidrule(lr){11-13}

Par.
& Bias & RMSE & ASE
& Bias & RMSE & ASE
& Bias & RMSE & ASE
& Bias & RMSE & ASE\\

\cmidrule(lr){2-4}
\cmidrule(lr){5-7}
\cmidrule(lr){8-10}
\cmidrule(lr){11-13}

$\beta_1$
& $-0.0029$ & 0.1398 & 0.1335
& 0.0005    & 0.0998 & 0.0961
& 0.0000    & 0.0697 & 0.0687
& $-0.0003$ & 0.0500 & 0.0488 \\

$\beta_2$
& 0.0081 & 0.2006 & 0.1920
& 0.0026 & 0.1421 & 0.1374
& 0.0024 & 0.0987 & 0.0976 
& 0.0007 & 0.0701 & 0.0693 \\

$\gamma_1$
& $-0.0328$ & 0.1781 & 0.1764
& $-0.0128$ & 0.1221 & 0.1212
& $-0.0096$ & 0.0848 & 0.0845
& $-0.0038$ & 0.0588 & 0.0592 \\

$\lambda_1$
& 0.0109    & 0.6168 & 0.6821
& $-0.0179$ & 0.4054 & 0.4130
& 0.0008    & 0.2727 & 0.2788
& $-0.0054$ & 0.1878 & 0.1913 \\

\bottomrule
\end{tabular}


\end{table}

\begin{table}[!ht]
\centering
\caption{ Bias, RMSE and ASE under Scenarios 7--10.}
\label{SimulationResults2}
\renewcommand{\arraystretch}{0.9}

\scriptsize
\setlength{\tabcolsep}{3pt}
\begin{tabular}{lrrrrrrrrrrrr}
\toprule

& \multicolumn{12}{c}{\textbf{Scenario 7}}\\
\cmidrule(lr){2-13}

& \multicolumn{3}{c}{$n=40$}
& \multicolumn{3}{c}{$n=80$}
& \multicolumn{3}{c}{$n=160$}
& \multicolumn{3}{c}{$n=320$}\\

\cmidrule(lr){2-4}
\cmidrule(lr){5-7}
\cmidrule(lr){8-10}
\cmidrule(lr){11-13}

Par.
& Bias & RMSE & ASE
& Bias & RMSE & ASE
& Bias & RMSE & ASE
& Bias & RMSE & ASE\\

\cmidrule(lr){2-4}
\cmidrule(lr){5-7}
\cmidrule(lr){8-10}
\cmidrule(lr){11-13}

$\beta_1$
& $-0.0007$ & 0.1200 & 0.1108
& $-0.0003$ & 0.0815 & 0.0795
& $-0.0001$ & 0.0572 & 0.0567
& $-0.0007$ & 0.0409 & 0.0403 \\

$\beta_2$
& $-0.0229$ & 0.3620 & 0.3359
& $-0.0070$ & 0.2479 & 0.2415
& $-0.0034$ & 0.1732 & 0.1724
& $-0.0031$ & 0.1226 & 0.1227 \\

$\gamma_1$
& $-0.0832$ & 0.2783 & 0.2578
& $-0.0380$ & 0.1828 & 0.1762
& $-0.0227$ & 0.1247 & 0.1223
& $-0.0109$ & 0.0858 & 0.0858 \\

$\gamma_2$
& $-0.0221$ & 0.4377 & 0.4056
& $-0.0129$ & 0.2770 & 0.2752
& $-0.0080$ & 0.1936 & 0.1905
& $-0.0060$ & 0.1339 & 0.1334 \\

$\lambda_1$
& 0.0498 & 0.5026 & 0.5373
& 0.0133 & 0.3381 & 0.3486
& 0.0174 & 0.2299 & 0.2353
& 0.0101 & 0.1635 & 0.1628 \\
\midrule

& \multicolumn{12}{c}{\textbf{Scenario 8}}\\
\cmidrule(lr){2-13}

& \multicolumn{3}{c}{$n=40$}
& \multicolumn{3}{c}{$n=80$}
& \multicolumn{3}{c}{$n=160$}
& \multicolumn{3}{c}{$n=320$}\\

\cmidrule(lr){2-4}
\cmidrule(lr){5-7}
\cmidrule(lr){8-10}
\cmidrule(lr){11-13}

Par.
& Bias & RMSE & ASE
& Bias & RMSE & ASE
& Bias & RMSE & ASE
& Bias & RMSE & ASE\\


\cmidrule(lr){2-4}
\cmidrule(lr){5-7}
\cmidrule(lr){8-10}
\cmidrule(lr){11-13}

$\beta_1$
& 0.0034 & 0.1620 & 0.1495
& 0.0017 & 0.1100 & 0.1072
& 0.0006 & 0.0776 & 0.0762
& $-0.0005$ & 0.0543 & 0.0541 \\

$\beta_2$
& 0.0098 & 0.4509 & 0.4165
& 0.0127 & 0.3116 & 0.2998
& 0.0090 & 0.2175 & 0.2134
& 0.0031 & 0.1516 & 0.1515 \\

$\gamma_1$
& $-0.0610$ & 0.2447 & 0.2234
& $-0.0286$ & 0.1593 & 0.1537 
& $-0.0156$ & 0.1082 & 0.1070
& $-0.0071$ & 0.0756 & 0.0752 \\

$\gamma_2$
& $-0.0129$ & 0.4396 & 0.4069
& $-0.0060$ & 0.2820 & 0.2770
& $-0.0029$ & 0.1945 & 0.1919
& $-0.0031$ & 0.1363 & 0.1344 \\

$\lambda_1$
& $-0.0292$ & 0.4379 & 0.4570
& $-0.0268$ & 0.2967 & 0.2991
& $-0.0086$ & 0.2020 & 0.2037
& $-0.0041$ & 0.1407 & 0.1413 \\
\midrule

& \multicolumn{12}{c}{\textbf{Scenario 9}}\\
\cmidrule(lr){2-13}

& \multicolumn{3}{c}{$n=40$}
& \multicolumn{3}{c}{$n=80$}
& \multicolumn{3}{c}{$n=160$}
& \multicolumn{3}{c}{$n=320$}\\

\cmidrule(lr){2-4}
\cmidrule(lr){5-7}
\cmidrule(lr){8-10}
\cmidrule(lr){11-13}

Par.
& Bias & RMSE & ASE
& Bias & RMSE & ASE
& Bias & RMSE & ASE
& Bias & RMSE & ASE\\

\cmidrule(lr){2-4}
\cmidrule(lr){5-7}
\cmidrule(lr){8-10}
\cmidrule(lr){11-13}

$\beta_1$
& $-0.0061$ & 0.2287 & 0.1992 
& $-0.0057$ & 0.1543 & 0.1459
& $-0.0002$ & 0.1089 & 0.1050
& $-0.0010$ & 0.0762 & 0.0752 \\

$\beta_2$
& $-0.0085$ & 0.4386 & 0.3815 
&   0.0030  & 0.2960 & 0.2784
& 0.0002 & 0.2096 & 0.2005
& $-0.0015$ & 0.1430 & 0.1435 \\

$\beta_3$
& $-0.0054$ & 0.1987 & 0.1800 
&   0.0005  & 0.1374 & 0.1311
& $-0.0019$ & 0.0983 & 0.0943
& $-0.0009$ & 0.0687 & 0.0674 \\

$\gamma_1$
& $-0.2049$ & 0.4673 & 0.3825
& $-0.0865$ & 0.2744 & 0.2498
& $-0.0454$ & 0.1829 & 0.1707
& $-0.0215$ & 0.1202 & 0.1185 \\

$\gamma_2$
& 0.1032 & 0.5805 & 0.5687
& 0.0255 & 0.3733 & 0.3670
& 0.0158 & 0.2523 & 0.2491
& 0.0078 & 0.1712 & 0.1723 \\

$\gamma_3$
& 0.0391 & 0.3389 & 0.3052
& 0.0182 & 0.2119 & 0.2050
& 0.0075 & 0.1463 & 0.1412
& 0.0016 & 0.0988 & 0.0986 \\

$\lambda_1$
&   0.0651  & 1.1503 & 1.3950 
& $-0.0534$ & 0.8323 & 0.8821 
& $-0.0201$ & 0.5704 & 0.5680
& 0.0017 & 0.3756 & 0.3814 \\

$\lambda_2$
& 0.0523 & 1.5827 & 1.8826
& 0.1168 & 1.1036 & 1.1670 
& 0.0558 & 0.7533 & 0.7480
& 0.0132 & 0.4988 & 0.5011 \\
\midrule

& \multicolumn{12}{c}{\textbf{Scenario 10}}\\
\cmidrule(lr){2-13}

& \multicolumn{3}{c}{$n=40$}
& \multicolumn{3}{c}{$n=80$}
& \multicolumn{3}{c}{$n=160$}
& \multicolumn{3}{c}{$n=320$}\\

\cmidrule(lr){2-4}
\cmidrule(lr){5-7}
\cmidrule(lr){8-10}
\cmidrule(lr){11-13}

Par.
& Bias & RMSE & ASE
& Bias & RMSE & ASE
& Bias & RMSE & ASE
& Bias & RMSE & ASE\\


\cmidrule(lr){2-4}
\cmidrule(lr){5-7}
\cmidrule(lr){8-10}
\cmidrule(lr){11-13}

$\beta_1$
& $-0.0109$ & 0.2730 & 0.2326
& $-0.0104$ &  0.1872 & 0.1726
& $-0.0026$ & 0.1296 & 0.1253
& $-0.0019$ & 0.0919 & 0.0898 \\

$\beta_2$
& 0.0067 & 0.6488 & 0.5597
& 0.0162 & 0.4468 & 0.4144
& 0.0047 & 0.3128 & 0.3002
& 0.0008 & 0.2186 & 0.2150 \\

$\beta_3$
& $-0.0286$ & 0.2820 & 0.2598
& $-0.0010$ & 0.1987 & 0.1883
& $-0.0043$ & 0.1384& 0.1348
& $-0.0045$ & 0.0972 & 0.0962 \\

$\gamma_1$
& $-0.2429$ & 0.5201 & 0.4202
& $-0.1141$ & 0.3105 & 0.2728
& $-0.0528$ & 0.2001 & 0.1861
& $-0.0281$ & 0.1347 & 0.1289 \\

$\gamma_2$
& 0.0988 & 0.6187 & 0.5788
& 0.0431 & 0.3829 & 0.3686
& 0.0172 & 0.2548 & 0.2491
& 0.0113 & 0.1734 & 0.1715 \\

$\gamma_3$
& 0.0667 & 0.3397 & 0.2875
& 0.0317 & 0.2094 & 0.1943
& 0.0120 & 0.1421 & 0.1347
& 0.0043 & 0.0947 & 0.0943 \\

$\lambda_1$
& 0.0993 & 0.7460 & 0.7832
& 0.0413 & 0.5004 & 0.5051
& 0.0154 & 0.3426 & 0.3405
& 0.0188 & 0.2348 & 0.2335 \\

$\lambda_2$
& $-0.0862$ & 0.9907 & 1.0428
& $-0.0458$ & 0.6480 & 0.6624
& $-0.0114$ & 0.4474 & 0.4430
& $-0.0189$ & 0.3050 & 0.3029 \\

\bottomrule
\end{tabular}


\end{table}

An important finding from the results is that the regression coefficients associated with the $\nu$ submodel generally exhibit larger ASEs and RMSEs than those associated with the $\mu$ and $\sigma$ submodels. This difference is more pronounced in Scenarios 9 and 10, where $\nu$ is modeled as a function of a covariate.
For instance, in Scenario 9 with $n=40$, the RMSEs of $\lambda_1$ and $\lambda_2$ are 1.1503 and 1.5827, respectively, whereas the RMSEs of the coefficients of $\mu$ range from 0.1987 to 0.4386. A similar pattern is observed in Scenario 10. These results suggest that the parameter $\nu$ and its associated regression coefficients are more challenging to estimate accurately than the parameters $\mu$ and $\sigma$.
The increasing complexity of the regression structure also affects the finite-sample performance of the estimators. Compared to Scenarios 1--6, in which only the $\mu$ submodel includes covariates, Scenarios 7 and 8 allow both $\mu$ and $\sigma$ to vary with covariates, resulting in somewhat larger RMSEs and ASEs for several regression coefficients. The largest estimation errors are observed in Scenarios 9 and 10, where all three model parameters depend on covariates. Nevertheless, even in these more demanding settings, the RMSEs and ASEs decrease as the sample size increases, and the biases approach zero.
Overall, the simulation results indicate that the proposed MLE procedure provides reliable parameter estimates across a wide range of regression structures. Although greater variability and finite-sample bias are observed for the parameter \(\nu\), particularly in more complex scenarios and with small sample sizes, these effects diminish substantially as the sample size increases.

Additionally, the behavior of the Wald hypothesis test was investigated for testing the null hypothesis $H_0$: $\theta_j = \theta_j^0$ against the two-sided alternative $H_1$:  $\theta_j \neq \theta_j^0$, where $\theta_j^0$ denotes the true value of the $j$th parameter under each simulation scenario. The Wald test rejects $H_0$ at the nominal level $\alpha$ when $|z_j| = |({\widehat{\theta}_j - \theta_j^0})/{\rm se}(\widehat{\theta}_j)|$ exceeds the $(1-\alpha/2)$ quantile of the standard normal distribution.
The empirical rejection rates were computed using a nominal significance level of 5\%. 

Table \ref{SimulationResults3} shows the empirical levels of the Wald tests for different sample sizes under Scenarios 1–-10.  Overall, the results show that the empirical rejection rates are close to the nominal level of 5\%, particularly as the sample size increases. 
This behavior is consistent with the asymptotic properties of the Wald tests and provides evidence that the tests adequately control the Type I error rate at the nominal 5\% level under the null hypothesis.
For Scenarios 1--6, the empirical levels are generally close to 0.05 across all sample sizes. Although some deviations from the nominal level are observed for $n=40$, these differences become smaller as the sample size increases. For example, in Scenario 1, the empirical levels for $\beta_1$ are 0.0638, 0.0602, 0.0492, and 0.0518 for $n=40$, $80$, $160$, and $320$, respectively. Similar behavior is observed for the other parameters and scenarios, with the empirical levels approaching 5\% as $n$ increases. A similar pattern is observed in Scenarios 7 and 8, where both the $\mu$ and $\sigma$ submodels include covariates. Despite the increased complexity of these models, the empirical levels remain reasonably close to the nominal 5\% level and tend to approach it as the sample size increases. For instance, in Scenario 7, the empirical level for $\beta_1$ decreases from 0.0794 for $n=40$ to 0.0554 for $n=320$. The largest departures from the nominal level occur in Scenarios 9 and 10 for the smaller sample sizes. In particular, the empirical levels for some coefficients in the $\mu$ and $\sigma$ submodels are somewhat higher than 5\% when $n=40$. For example, the empirical levels for $\beta_1$ and $\gamma_1$ in Scenario 10 are 0.1174 and 0.1000, respectively. However, these deviations diminish substantially as the sample size increases. At $n=320$, the corresponding empirical levels are 0.0608 and 0.0564, respectively. Thus, even under the most complex regression structures, the empirical Type I error rates approach the nominal 5\% level as the sample size increases.

Regarding the $\nu$ coefficients, the empirical levels are below the nominal 5\% level for the smallest sample sizes in several scenarios, particularly in Scenarios 2, 5, 7, 9, and 10. For example, in Scenario 9, the empirical levels for $\lambda_1$ and $\lambda_2$ are 0.0070 and 0.0090, respectively, when $n=40$. However, these rates increase toward the nominal level as the sample size increases, reaching 0.0444 and 0.0400, respectively, for $n=320$. A similar pattern is observed in Scenario 10. Overall, the results provide strong empirical evidence that the Wald tests achieve the intended 5\% significance level asymptotically. Although moderate deviations from the nominal level may occur with small samples, the empirical rejection rates consistently approach 5\% as $n$ increases. This pattern is observed across the various regression structures and parameter configurations considered, including the most complex scenarios where all three parameters of the BSG model depend on covariates.

\begin{table}[!ht]
\centering
\caption{Empirical null levels of Wald tests under Scenarios 1--10 at the 5\% nominal level.}
\label{SimulationResults3}

\scriptsize
\renewcommand{\arraystretch}{0.85}
\setlength{\tabcolsep}{3pt}

\begin{tabular}{lcccc|cccc}
\toprule
& \multicolumn{4}{c|}{\textbf{Scenario 1}}
& \multicolumn{4}{c}{\textbf{Scenario 2}}\\
\cmidrule(lr){2-5}\cmidrule(lr){6-9}
Par. & $n=40$ & $n=80$ & $n=160$ & $n=320$ & $n=40$ & $n=80$ & $n=160$ & $n=320$\\
\midrule
$\beta_1$& 0.0638  & 0.0602 & 0.0492 & 0.0518 & 0.0694 & 0.0578 & 0.0518 & 0.0504\\
$\beta_2$& 0.0676  & 0.0548 & 0.0562 & 0.0488 & 0.0682 & 0.0552 & 0.0554 & 0.0530\\
$\gamma_1$& 0.0692 & 0.0548 & 0.0608 & 0.0528 & 0.0628 & 0.0526 & 0.0548 & 0.0500\\
$\lambda_1$&0.0586 & 0.0498 & 0.0514 & 0.0430 & 0.0016 & 0.0246 & 0.0390 & 0.0378\\
\midrule

& \multicolumn{4}{c|}{\textbf{Scenario 3}}
& \multicolumn{4}{c}{\textbf{Scenario 4}}\\
\cmidrule(lr){2-5}\cmidrule(lr){6-9}
Par. & $n=40$ & $n=80$ & $n=160$ & $n=320$ & $n=40$ & $n=80$ & $n=160$ & $n=320$\\
\midrule
$\beta_1$&  0.0666 & 0.0608 & 0.0506 & 0.0508 & 0.0660 & 0.0616 & 0.0480 & 0.0512\\
$\beta_2$&  0.0644 & 0.0612 & 0.0524 & 0.0528 & 0.0674 & 0.0540 & 0.0564 & 0.0476\\
$\gamma_1$& 0.0486 & 0.0548 & 0.0552 & 0.0446 & 0.0698 & 0.0562 & 0.0606 & 0.0494\\
$\lambda_1$&0.0538 & 0.0592 & 0.0432 & 0.0434 & 0.0618 & 0.0450 & 0.0490 & 0.0404\\
\midrule

& \multicolumn{4}{c|}{\textbf{Scenario 5}}
& \multicolumn{4}{c}{\textbf{Scenario 6}}\\
\cmidrule(lr){2-5}\cmidrule(lr){6-9}
Par. & 40 & 80 & 160 & 320 & 40 & 80 & 160 & 320\\
\midrule
$\beta_1$&  0.0674 & 0.0622 & 0.0536 & 0.0546 & 0.0706 & 0.0614 & 0.0512 & 0.0552\\
$\beta_2$&  0.0668 & 0.0582 & 0.0560 & 0.0532 & 0.0650 & 0.0624 & 0.0536 & 0.0536\\
$\gamma_1$& 0.0666 & 0.0540 & 0.0526 & 0.0510 & 0.0546 & 0.0526 & 0.0510 & 0.0478\\
$\lambda_1$&0.0124 & 0.0370 & 0.0438 & 0.0434 & 0.0538 & 0.0546 & 0.0474 & 0.0472\\
\midrule

& \multicolumn{4}{c|}{\textbf{Scenario 7}}
& \multicolumn{4}{c}{\textbf{Scenario 8}}\\
\cmidrule(lr){2-5}\cmidrule(lr){6-9}
Par. & $n=40$ & $n=80$ & $n=160$ & $n=320$ & $n=40$ & $n=80$ & $n=160$ & $n=320$\\
\midrule
$\beta_1$&  0.0794 & 0.0648 & 0.0538 & 0.0554 & 0.0804 & 0.0558 & 0.0546 & 0.0540\\
$\beta_2$&  0.0752 & 0.0592 & 0.0540 & 0.0492 & 0.0774 & 0.0600 & 0.0570 & 0.0504\\
$\gamma_1$& 0.0652 & 0.0594 & 0.0554 & 0.0480 & 0.0716 & 0.0588 & 0.0566 & 0.0554 \\
$\gamma_2$& 0.0688 & 0.0524 & 0.0516 & 0.0490 & 0.0726 & 0.0556 & 0.0532 & 0.0534\\
$\lambda_1$&0.0170 & 0.0336 & 0.0410 & 0.0460 & 0.0272 & 0.0440 & 0.0482 & 0.0458\\
\midrule

& \multicolumn{4}{c|}{\textbf{Scenario 9}}
& \multicolumn{4}{c}{\textbf{Scenario 10}}\\
\cmidrule(lr){2-5}\cmidrule(lr){6-9}
Par. & $n=40$ & $n=80$ & $n=160$ & $n=320$ & $n=40$ & $n=80$ & $n=160$ & $n=320$\\
\midrule
$\beta_1$&  0.1008 & 0.0666 & 0.0672 & 0.0592 & 0.1174 & 0.0824 & 0.0618 & 0.0608\\
$\beta_2$&  0.1014 & 0.0702 & 0.0592 & 0.0486 & 0.1152 & 0.0830 & 0.0686 & 0.0554\\
$\beta_3$&  0.0884 & 0.0712 & 0.0664 & 0.0560 & 0.0768 & 0.0698 & 0.0566 & 0.0518\\
$\gamma_1$& 0.0998 & 0.0690 & 0.0646 & 0.0548 & 0.1000 & 0.0768 & 0.0662 & 0.0564\\
$\gamma_2$& 0.0552 & 0.0506 & 0.0542 & 0.0490 & 0.0562 & 0.0516 & 0.0540 & 0.0504\\
$\gamma_3$& 0.0754 & 0.0584 & 0.0538 & 0.0520 & 0.0972 & 0.0726 & 0.0646 & 0.0454\\
$\lambda_1$&0.0070 & 0.0230 & 0.0448 & 0.0444 & 0.0266 & 0.0452 & 0.0522 & 0.0524\\
$\lambda_2$&0.0090 & 0.0228 & 0.0444 & 0.0400 & 0.0298 & 0.0460 & 0.0488 & 0.0504\\
\bottomrule
\end{tabular}

\end{table}

\section{Real data applications}\label{sec6}

\subsection{Cheese data}

In this section, we present the first real data application based on a study investigating the relationship between the chemical composition of cheddar cheese and its taste. The data set, named {\tt cheese}, is available in the {\tt GLMsData} package \citep{GLMsDatapackage} for the {\tt R} software. Our objective is to model the cheese taste score ({\tt Taste}) as a function of lactic acid concentration ({\tt Lactic}) and the natural logarithm of the H2S concentration, {\tt log(H2S)}.

\begin{figure}[!htb]
	\centering
	\subfigure{\includegraphics[width=5.5cm,height=5cm]{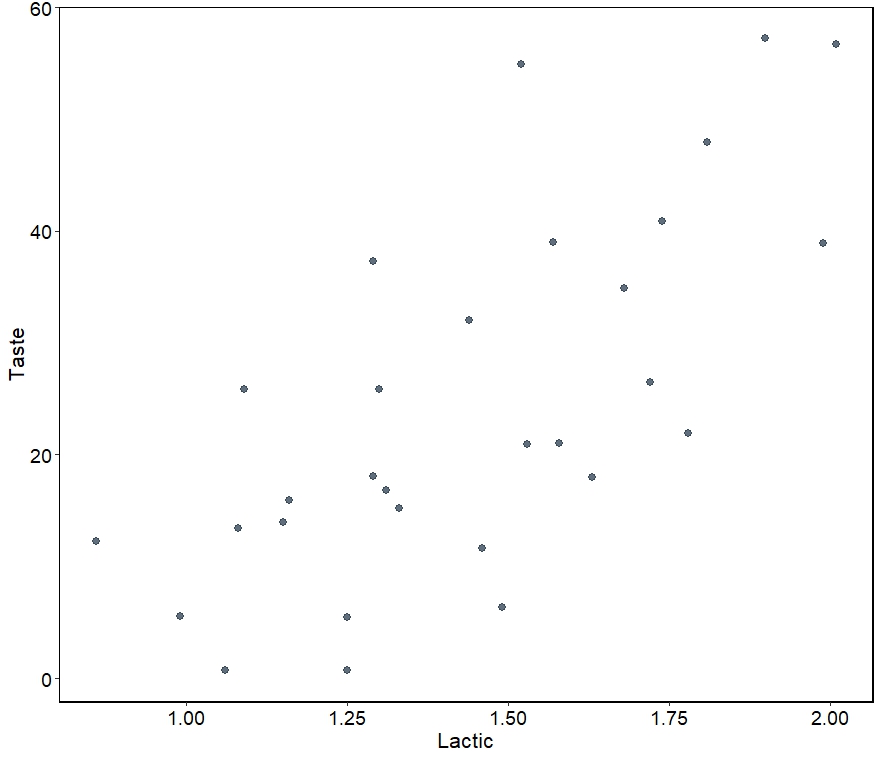}}
	\subfigure{\includegraphics[width=5.5cm,height=5cm]{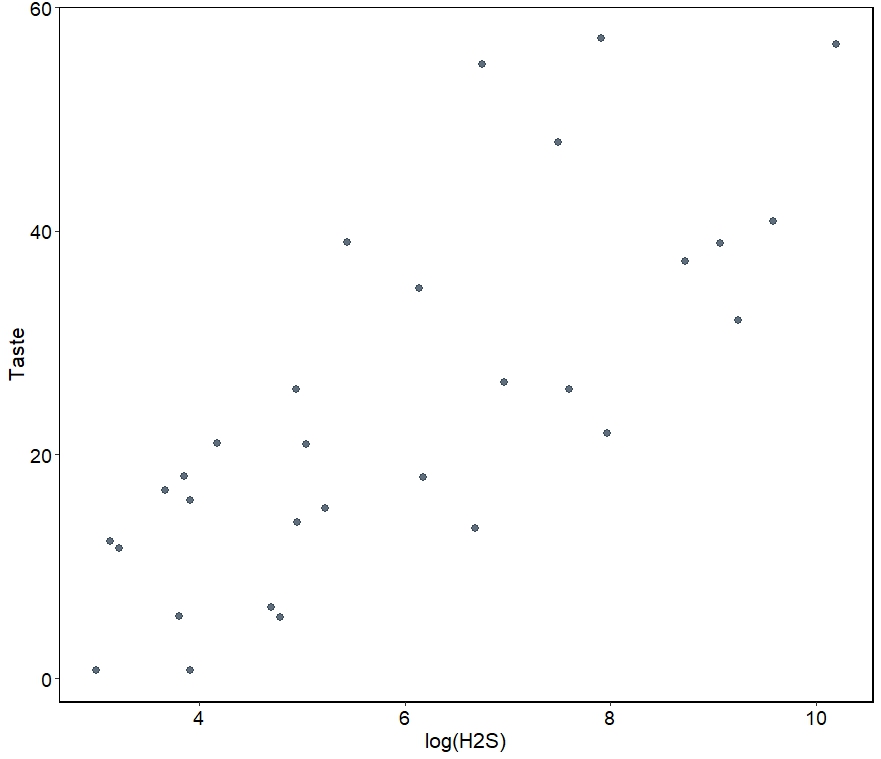}}
    \subfigure{\includegraphics[width=5.5cm,height=5cm]{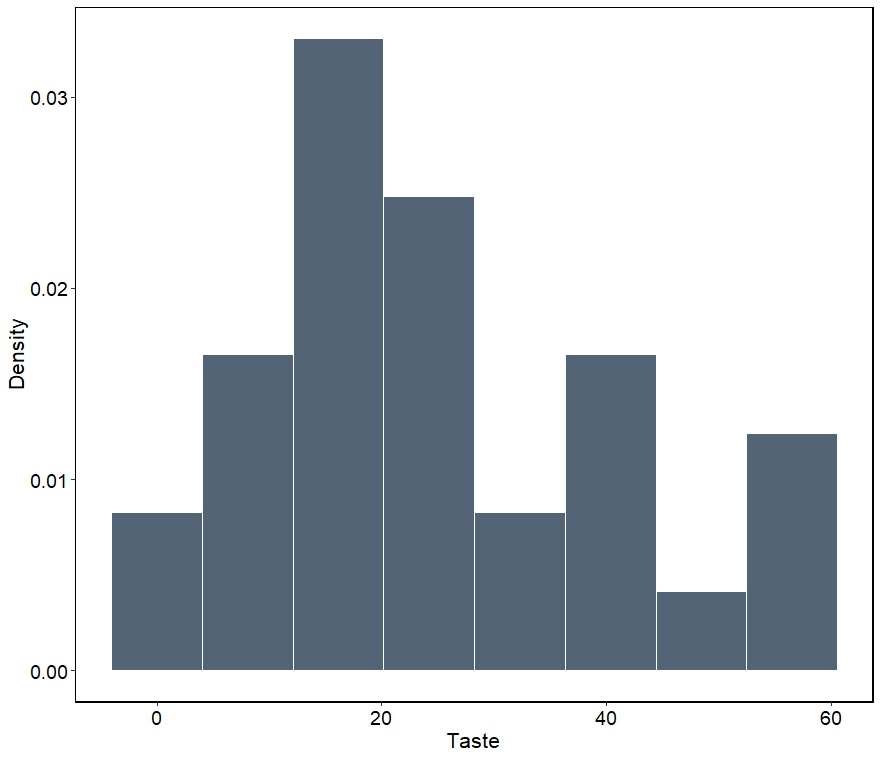}}
    \caption{Histogram of Taste and scatterplots against the covariates.}
	\label{tasteplots}
\end{figure}

Figure \ref{tasteplots} displays the histogram of {\tt Taste} along with its scatterplots against the covariates {\tt Lactic} and {\tt log(H2S)}. The plots suggest a positive relationship between the taste score and both covariates, suggesting that cheeses with higher lactic acid concentrations and higher $\log(\mathrm{H}_2\mathrm{S})$ values tend to have higher taste scores. Additionally, the marginal distribution of {\tt Taste} appears positively skewed, suggesting that an asymmetric distribution may be suitable for describing the response variable.

To demonstrate the advantages of the proposed GBS model over the BS model, we also fitted the BS model as a special case of the proposed implementation within the GAMLSS framework.

The regression models considered here specify the parameter $\mu_i$, which represents the median of the $i$th response, as
\begin{eqnarray*}
\log(\mu_i) = \beta_1 + \beta_2{\rm \log(H2S)}_i + \beta_3 {\rm  Lactic}_i,
\end{eqnarray*}
where $i=1, \ldots, 30$,  ${\rm  Lactic}_i$ and ${\rm \log(H2S)}_i$ are the lactic acid concentration and the natural logarithm of the H2S concentration of the $i$th  cheese, respectively.
 The parameter $\sigma$ in both models is assumed to be constant (i.e., not influenced by the covariates) and is expressed as $\log(\sigma)=\gamma_1$. For the GBS model, the  parameter $\nu$ is specified through the logit link and also assumed to be constant expressed by $\log(\nu/(1-\nu))=\lambda_1$. 

The AIC and BIC values for the GBS model are 228.07 and 235.07, respectively, while those for the BS model are 252.27 and 257.88, respectively. The lower values obtained for the GBS model under both criteria indicate a superior fit to the cheese data, suggesting that the additional flexibility introduced by the parameter $\nu$ is supported by the data.

Table \ref{estimates} shows the estimates, standard errors, $z$-statistics, and $p$-values of the test of nullity of coefficients for the GBS and BS regression models fitted to the cheese data. 
 At the 5\% significance level, the two models lead to different inferential conclusions. For the BS model, none of the regression coefficients in the $\mu$ submodel are statistically significant, suggesting that neither lactic acid concentration nor the natural logarithm of the H2S concentration provides sufficient evidence of an association with the median taste score. In contrast, under the GBS model, all coefficients in the $\mu$ submodel are statistically significant at the 5\% level, indicating that both covariates are significantly associated with the conditional median of the taste score. Furthermore, the estimate of $\nu$ is \(\widehat{\nu} =\exp(-2.313)= 0.099\), which is well below 0.5, the value corresponding to the BS model.  Additionally,  the coefficient $\lambda_1$ is statistically significant, with a $p$-value of 0.0075. Therefore, the null hypothesis $H_0$: $\lambda_1=0$ is rejected at the 5\% significance level. Since the logit link implies that $\lambda_1=0$ is equivalent to $\nu=0.5$, this result provides evidence that \( \nu \neq 0.5 \). Consequently, based on the results of this Wald test, the cheese data provide statistical support for including the additional  parameter $\nu$, indicating that the GBS model offers a more flexible fit than the corresponding BS model.
\begin{table}[!htb]
	\centering
	\caption{Estimates, standard errors,  $z$-stat and $p$-values for the fitted  regression models -- cheese data.}\vspace{0.2cm}
	\label{estimates}
	\scalefont{0.85}
	\def\arraystretch{1} \begin{tabular}{rrrrrrrrrr}\hline
& \multicolumn{4}{c}{GBS model}                                   & \multicolumn{4}{c}{BS model } \\  \cmidrule{2-5}\cmidrule{7-10}
                         &  Estimate & Std. error & $z$-stat &   $p$-value  && Estimate   &Std. error &$z$-stat   &$p$-value \\ \cmidrule{2-5}\cmidrule{7-10}
\text{$\mu$ submodel}    &&&&\\
		$\beta_1$        & 0.9208 & 0.4422 & 2.0820 & 0.0477    && $-0.5043$ & 1.1694 & $-0.4310$ & 0.6698    \\
		$\beta_2$        & 0.1540 & 0.0511 & 3.0110 & 0.0059    && 0.3089    & 0.1645  &  1.8780  & 0.0716    \\
		$\beta_3$        & 0.8350 & 0.3513 & 2.3770 & 0.0254    && 0.9810    & 0.7237  &  1.3560  & 0.1869  \\\cmidrule{2-5}\cmidrule{7-10}
        {\it $\sigma$} \text{submodel}    &&&\\
        $\gamma_1$       & 0.5028 & 0.1938 & 2.5940 & 0.0156    && $-0.1227$ & 0.2399  & $-0.5110$& 0.6130 \\
  \cmidrule{2-5}\cmidrule{7-10}\\
  {\it $\nu$} \text{submodel}    &&&\\
        $\lambda_1$      &$-2.3130$ & 0.7950 & $-2.9090$ & 0.0075   && --   & --  &   --   & --     \\\cmidrule{2-5}\cmidrule{7-10}\\\hline
	\end{tabular}
\end{table}

\begin{figure}[!htb]
	\centering
	\subfigure[Worm plot for GBS model]{\includegraphics[width=5cm,height=3.7cm]{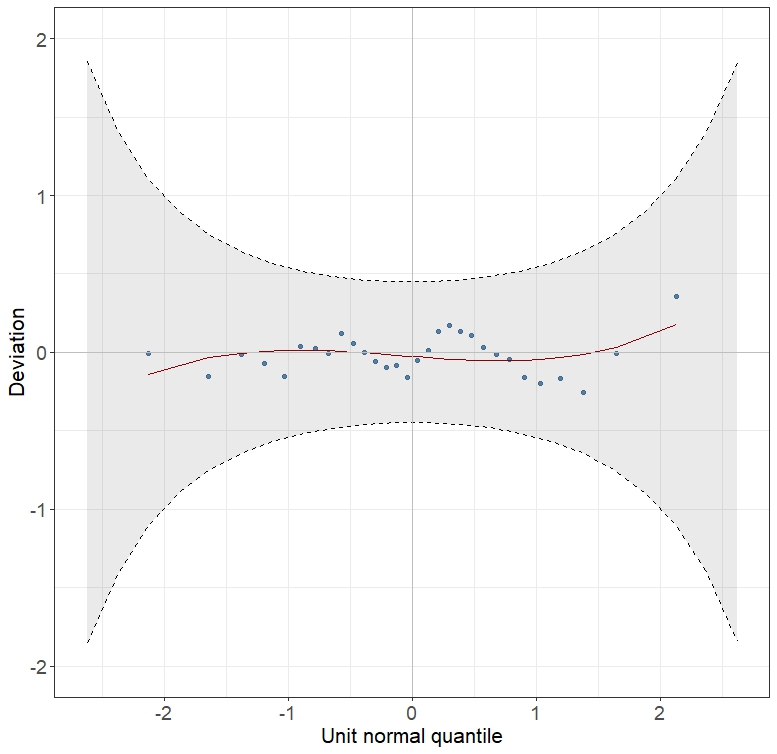}}
	\qquad
    \subfigure[Worm plot for BS model]{\includegraphics[width=5cm,height=3.7cm]{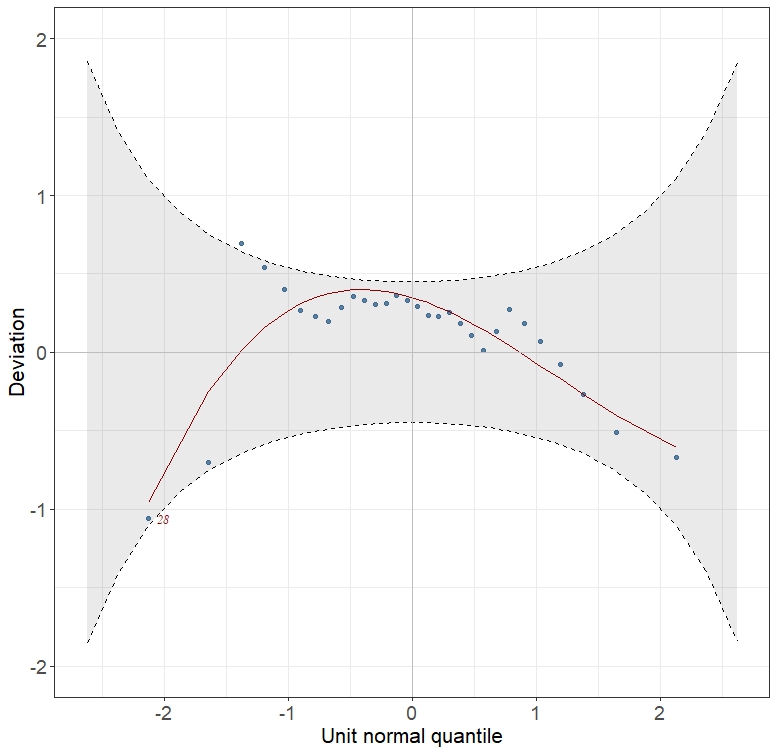}}\\
    \subfigure[Q-Q boxplot for GBS model]{\includegraphics[width=5cm,height=3.7cm]{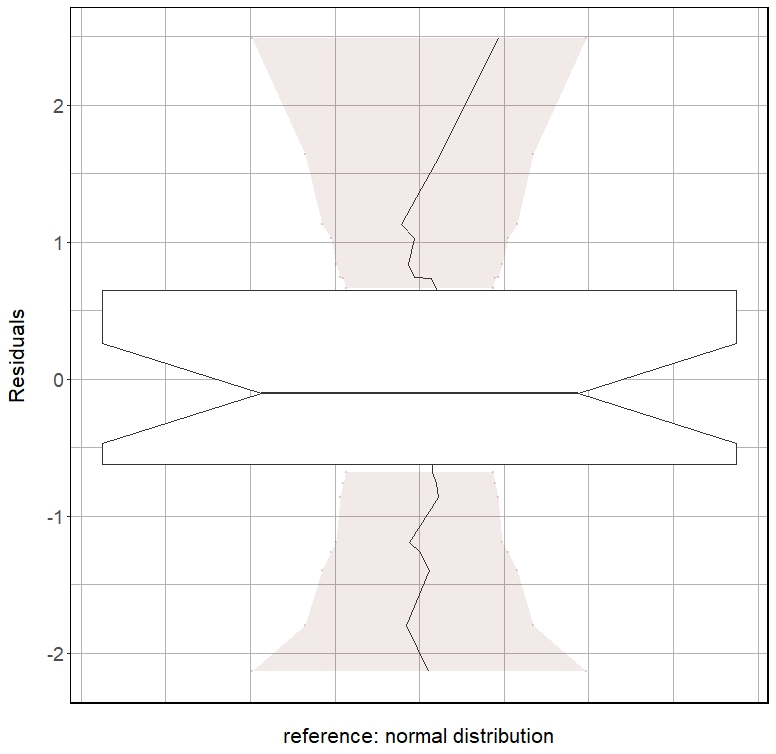}}
    \qquad
	\subfigure[Q-Q boxplot for BS model]{\includegraphics[width=5cm,height=3.7cm]{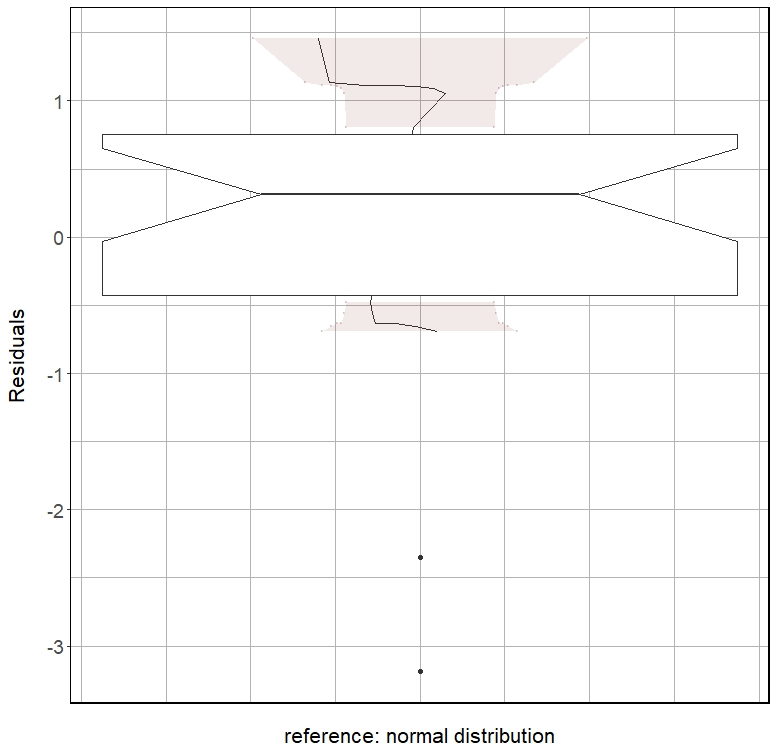}}\\
    \subfigure[Q-Q plot for GBS model]{\includegraphics[width=5cm,height=3.7cm]{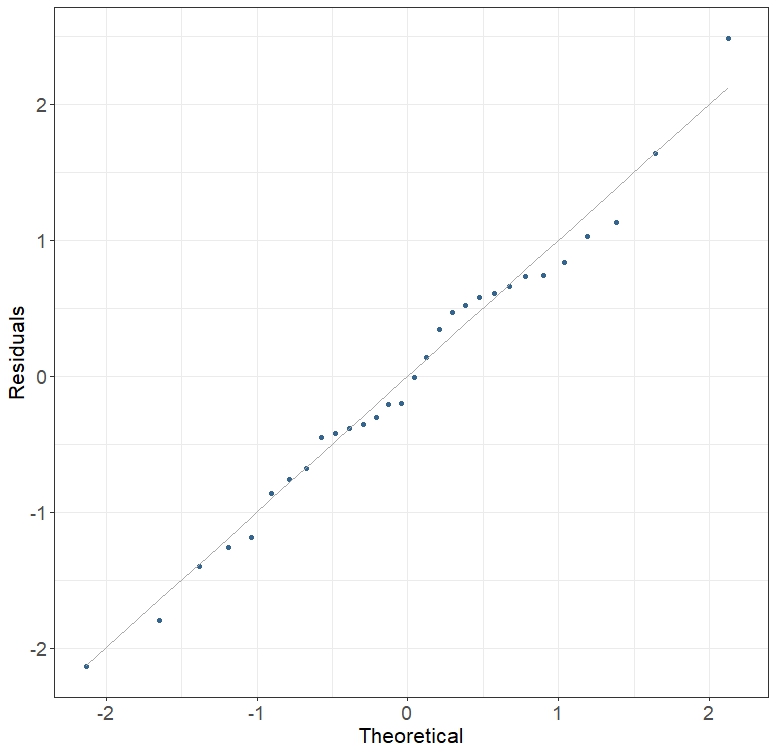}}
    \qquad
	\subfigure[Q-Q plot for BS model]{\includegraphics[width=5cm,height=3.7cm]{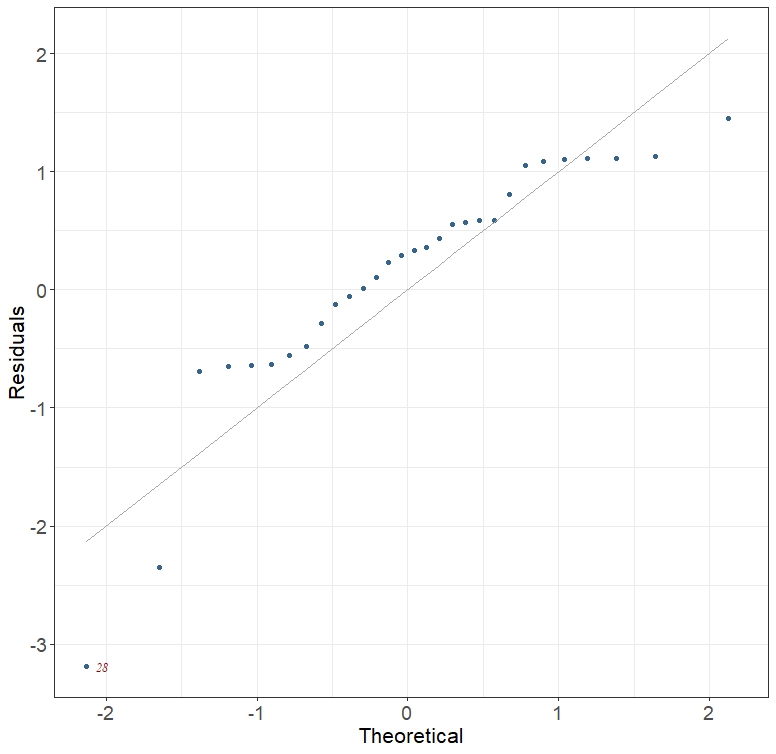}}\\
    \subfigure[Histogram of residuals with density estimator for GBS model]{\includegraphics[width=5cm,height=3.7cm]{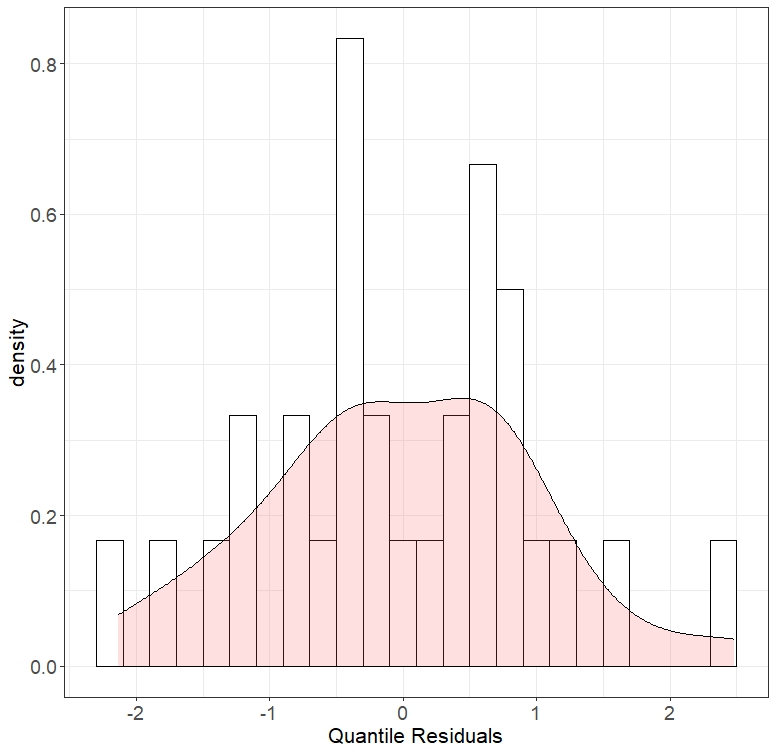}}
    \qquad
	\subfigure[Histogram of residuals with density estimator for BS model]{\includegraphics[width=5cm,height=3.7cm]{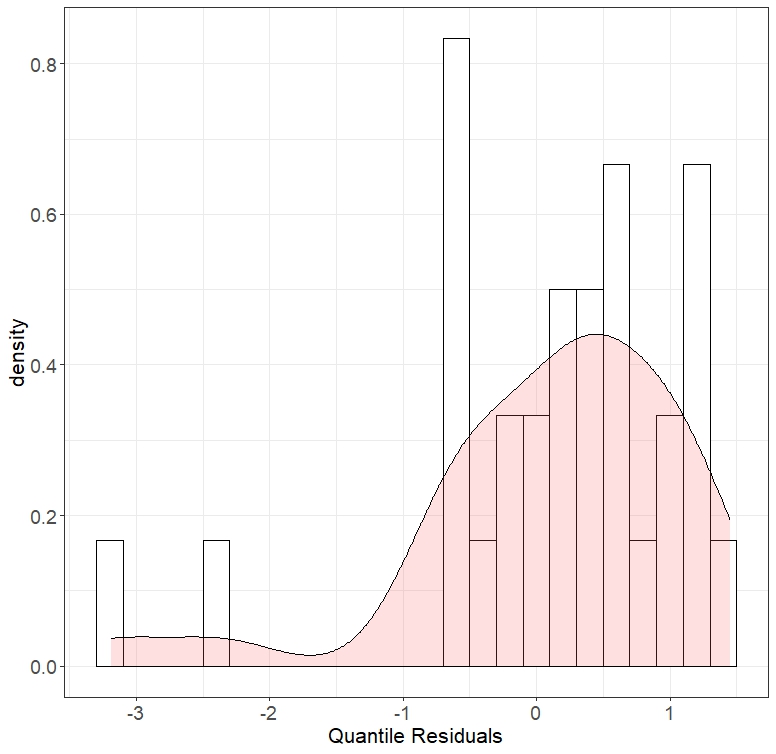}}
	\caption{Diagnostic plots based on the quantile residuals under the GBS and BS models -- cheese data. } 
	\label{diagplots_1}
\end{figure}

To assess the goodness of the fitted regression models, we consider worm plot,  Q--Q boxplots, Q--Q plots, and the histogram with density estimator of the quantile residuals, as shown in  Figure \ref{diagplots_1}.

Figure \ref{diagplots_1} reveals several indications of an inadequate fit for the BS model. In particular, the worm plot shows substantial deviations of the points from the central horizontal reference line, suggesting that the quantile residuals do not behave as expected under a correctly specified model. The pronounced inverted-U shape provides further evidence of systematic deviations and indicates that the BS specification may lack sufficient flexibility to capture the asymmetry present in the conditional response distribution. Similar evidence is observed in the Q-Q boxplot, where the residuals exhibit clear departures from the expected reference distribution, with the residual median considerably displaced from zero, and the overall pattern indicates a marked degree of asymmetry. This departure from symmetry is also evident in both the Q-Q plot and the histogram with a kernel density estimate. 
In contrast, the diagnostic plots based on the GBS fit indicate a good fit to the data. The worm plot for the GBS model shows a substantially improved fit compared to the BS fit, with points remaining close to the central horizontal line and no apparent systematic pattern. Similar conclusions are supported by the Q-Q boxplot, Q-Q plot, and the histogram with a kernel density estimate of the quantile residuals. 
Overall, the diagnostic plots provide additional evidence of the superior fit of the GBS model.

Considering the residual diagnostics, along with the AIC and BIC values and the results of the Wald test, we select the GBS model as the most appropriate compared to the BS model for the cheese data.

Under the GBS regression model, the results can be interpreted as follows.  With the lactic acid concentration kept fixed,  a one-unit increase in log(H2S) is associated with a 16.65\% increase in the median cheese taste score, since $\exp(0.1540)=1.1665$. Conversely, with log(H2S) fixed, a 0.5-unit increase in lactic acid concentration is associated with a 51.82\% increase in the median cheese taste score, since $\exp(0.8350\times0.5)=1.5182$.

\subsection{Rent data}

As a second application to real data, we consider the 1999 Munich rent survey, which consists of 3082 observations. The data set, denoted by {\tt rent99}, is available in the {\tt gamlss.data} package \citep{gamlssdata} for the {\tt R} software. These data have been modeled by \cite{de2018gaussian} , \cite{rigby2019distributions}, \cite{dutta2026} using different flexible distributions. 
The goal is to model the monthly net rent, measured in euros (\texttt{rent}), of flats in the city of Munich as a function of living area, measured in square meters (\texttt{area}); year of construction (\texttt{yearc}); location quality (\texttt{location}), categorized as average (1), good (2), or top (3); bathroom quality (\texttt{bath}), indicating whether the bathroom facilities are standard (0) or premium (1); and kitchen quality (\texttt{kitchen}), a binary indicator with 0 representing a standard kitchen and 1 representing a premium kitchen.
Following the approach adopted by the cited authors, the variable \texttt{yearc} is treated as a quantitative covariate.

\begin{figure}[!htb]
	\centering
	\subfigure{\includegraphics[width=4.2cm,height=4.5cm]{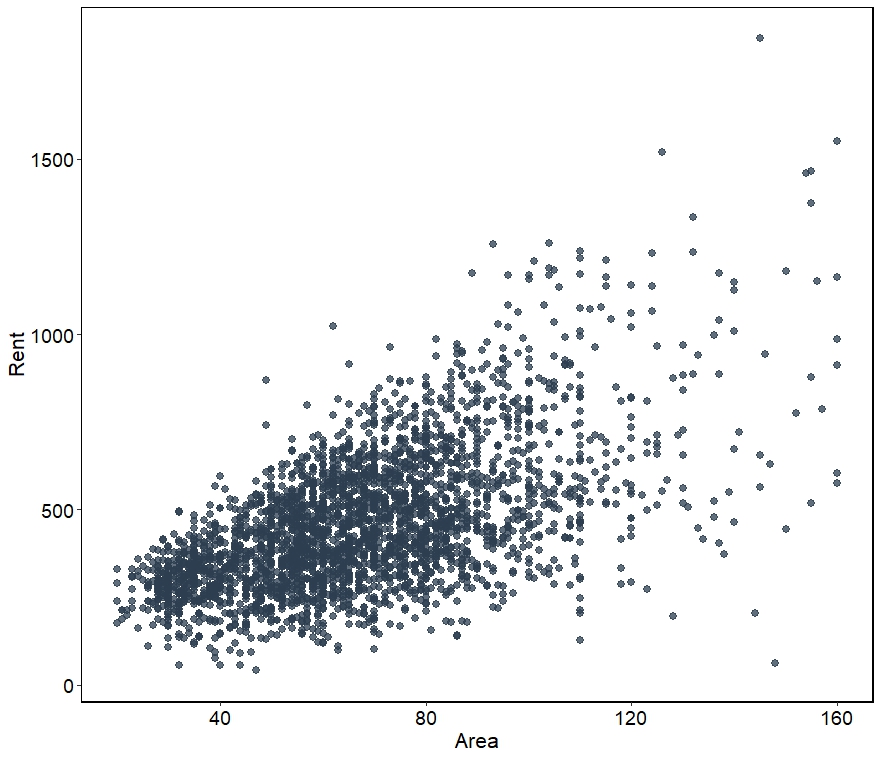}}
	\subfigure{\includegraphics[width=4.2cm,height=4.5cm]{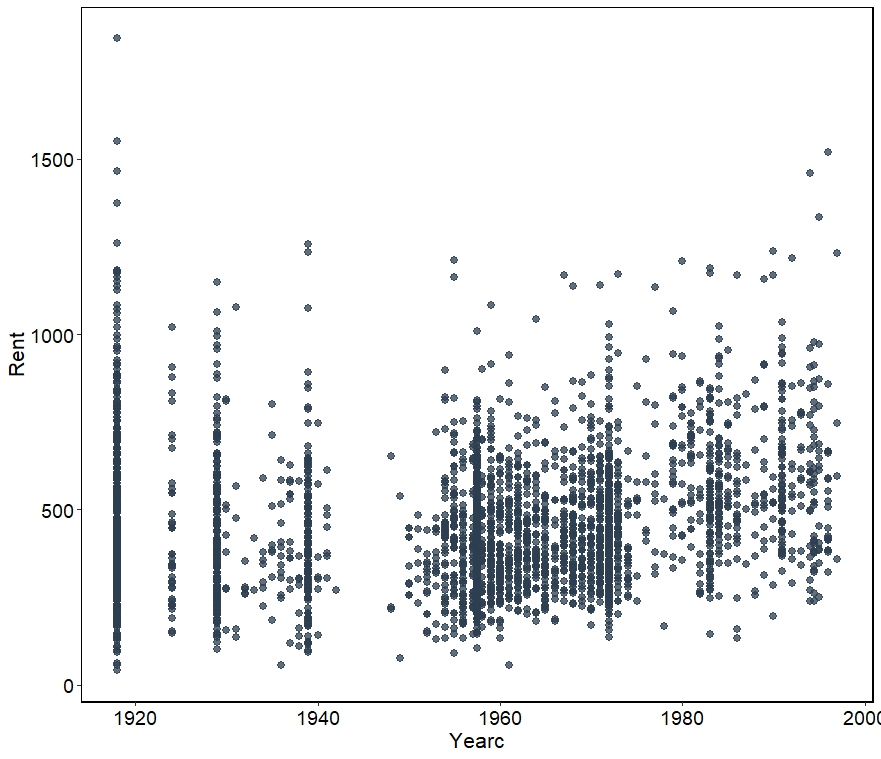}}
    \subfigure{\includegraphics[width=4.2cm,height=4.5cm]{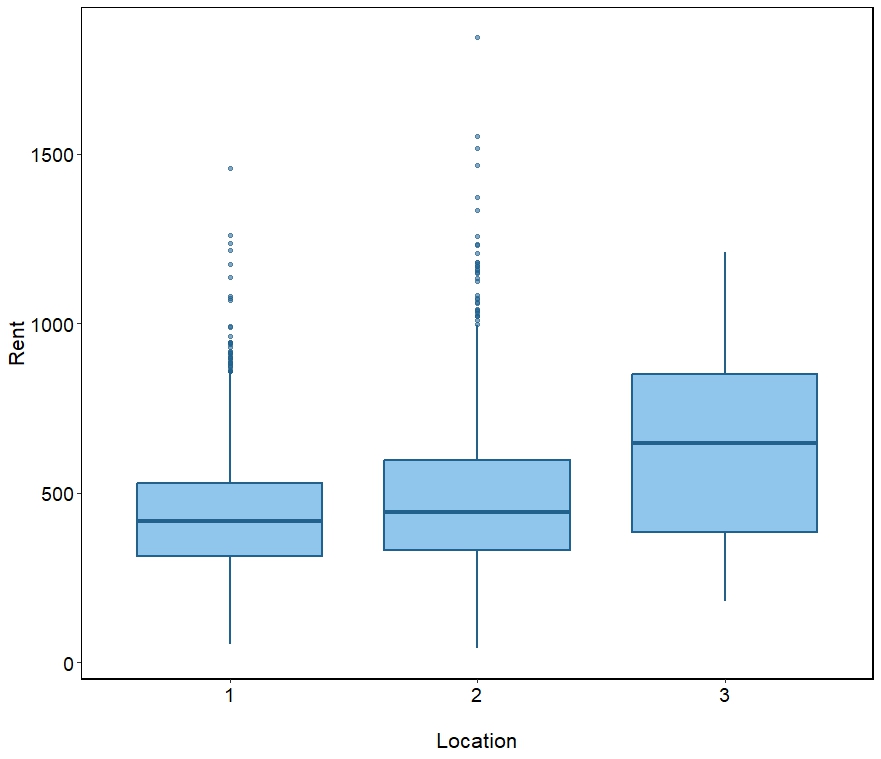}}\\
	\subfigure{\includegraphics[width=4.2cm,height=4.5cm]{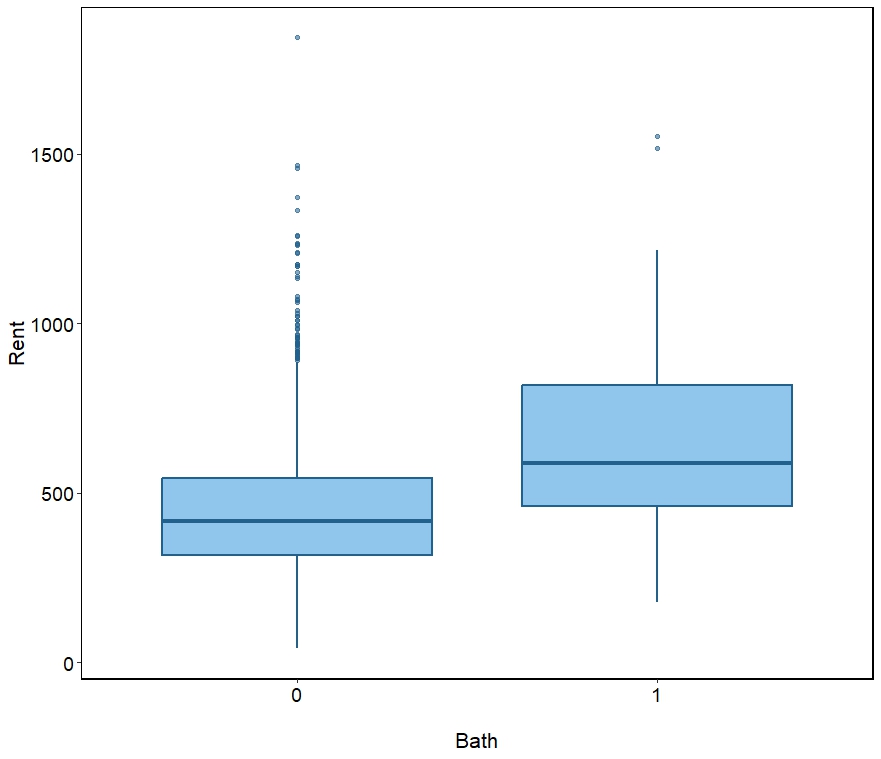}}
    \subfigure{\includegraphics[width=4.2cm,height=4.5cm]{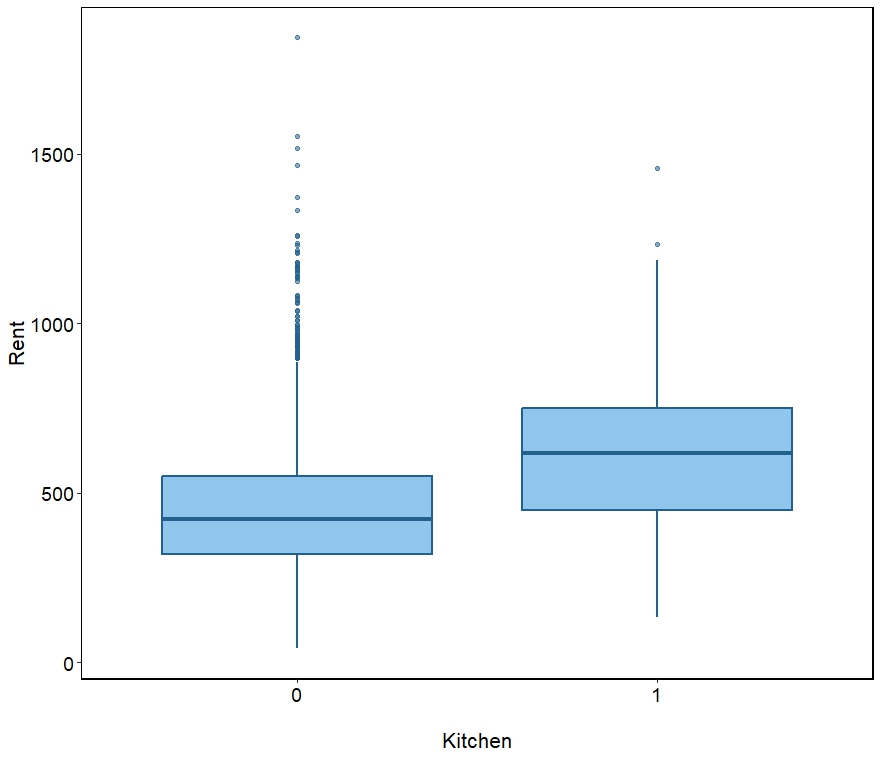}}
    \subfigure{\includegraphics[width=4.2cm,height=4.5cm]{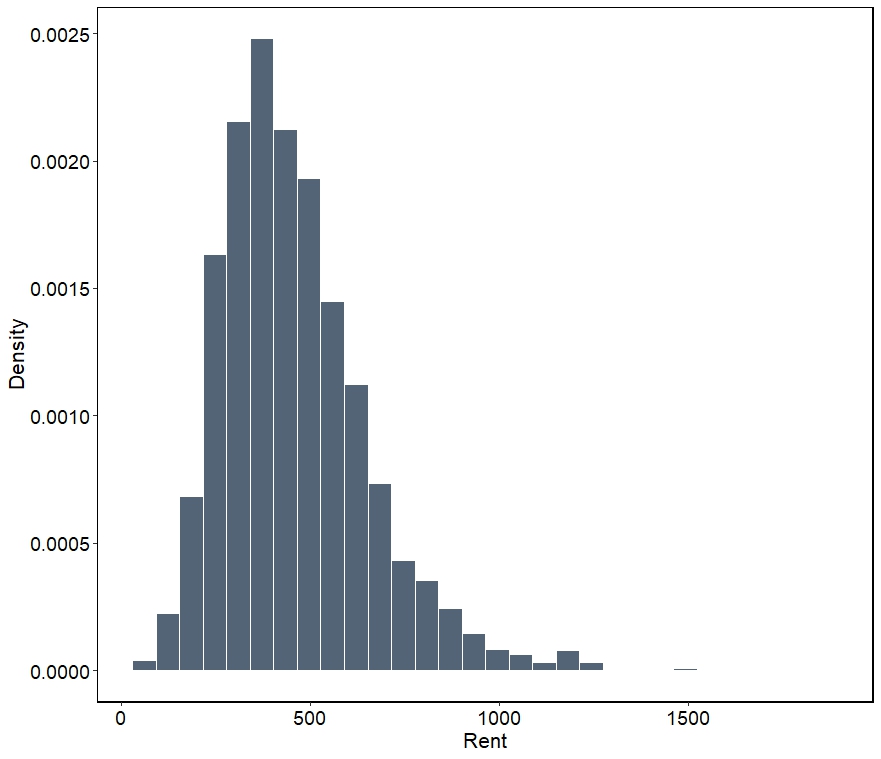}}
	\caption{Histogram of monthly net rent and scatterplots against the covariates. }
	\label{rentplots}
\end{figure}

Figure \ref{rentplots} presents the marginal distribution of monthly net rent and its relationship with the covariates. A positive relationship is observed between living area and rent, with larger flats generally commanding higher rental prices. Additionally, the variability in rents increases as the size of the flat grows, indicating greater heterogeneity in rental values for larger flats.
The year of construction appears to have a weaker relationship with rent. However, the plots indicate a modest increase in rental prices for flats built after approximately 1960. Additionally, flats in more desirable locations tend to have higher rents, accompanied by greater variability in rental values. Similar patterns are observed for bathroom and kitchen quality, as flats with premium facilities have higher monthly rents. The histogram of rent exhibits positive skewness, characterized by a right tail.

The GBS and BS regression models specify the parameter $\mu_i$ as
\begin{eqnarray*}
\log(\mu_i) = \beta_1 + \beta_2{\rm area}_i + \beta_3{\rm yearc}_i +
\beta_4{\rm loc2}_i + \beta_5{\rm loc3}_i + \beta_6{\rm bath}_i + \beta_7{\rm kitchen}_i ,
\end{eqnarray*}
where $i=1, \ldots, 3082$. Here,  ${\rm area}_i$  and ${\rm yearc}_i$  denote the living area and year of construction of the of the $i$th flat, respectively. The dummy variables ${\rm loc2}_i$  and  ${\rm loc3}_i$ indicate whether the $i$th flat is located in a good or top location, respectively, with the average location category serving as the reference level.  In addition,  ${\rm bath}_i$ and ${\rm kitchen}_i$  are binary indicators that take the value 1 if the $i$th flat has premium bathroom facilities and a premium kitchen, respectively, and 0 otherwise.
 The parameter $\sigma$ is specified by 
$\log(\sigma_i) = \gamma_1 + \gamma_2{\rm area}_i + \gamma_3{\rm loc2}_i + \gamma_4{\rm loc3}_i$. For the GBS model, the  parameter $\nu$ is expressed as $\log(\nu/(1-\nu))=\lambda_1$, i.e, it is specified through the logit link and assumed to be constant. 


Table \ref{estimates2} shows the estimates, standard errors, $z$-statistics, and $p$-values of the test of nullity of coefficients for the GBS and BS regression models fitted to the rent data. 
 At the 5\% significance level, all regression coefficients  are statistically significant, suggesting that the included covariates are relevant for explaining the median monthly net rent. In particular, the estimate of $\nu$ is \(\widehat{\nu} = \exp(-2.1870) = 0.112\), which is well below 0.5, the value corresponding to the BS model.  Furthermore,  the coefficient $\lambda_1$ of the GBS model is statistically significant, with a $p$-value less than 0.0001. Therefore, the null hypothesis $H_0$: $\lambda_1=0$ is rejected at the 5\% significance level, providing evidence that $\nu\neq0.5$. Consequently, this indicates that the fits obtained by the GBS and BS models differ. 
\begin{table}[!htb]
	\centering
	\caption{Estimates, standard errors,  $z$-stat and $p$-values for the fitted  regression models -- rent data.}\vspace{0.2cm}
	\label{estimates2}
	\scalefont{0.75}
	\def\arraystretch{1} \begin{tabular}{rrrrrrrrrr}\hline
& \multicolumn{4}{c}{GBS model}                                   & \multicolumn{4}{c}{BS model } \\  \cmidrule{2-5}\cmidrule{7-10}
                         &  Estimate & Std. error & $z$-stat &   $p$-value  && Estimate   &Std. error &$z$-stat   &$p$-value \\ \cmidrule{2-5}\cmidrule{7-10}
\text{$\mu$ submodel}    &&&&\\
		$\beta_1$        & $-4.9823$ & 0.4412 & $-11.293$ & $<0.0001$  && $-7.0562$ & 0.5154 & $-13.6920$ & $<0.0001$  \\
		$\beta_2$        &   0.0110  & 0.0003 &   41.985  & $<0.0001$  &&   0.0109  & 0.0002 &   48.1520  & $<0.0001$ \\
        $\beta_3$        &   0.0053  & 0.0002 &   23.478  & $<0.0001$  &&   0.0063  & 0.0003 &   23.9830  & $<0.0001$ \\
        $\beta_4$        &   0.0873  & 0.0113 &   7.7440  & $<0.0001$  &&   0.0810  & 0.0124 &   6.5580   & $<0.0001$  \\
        $\beta_5$        &   0.1998  & 0.0402 &   4.9680  & $<0.0001$  &&   0.1818  & 0.0441 &   4.1240   & $<0.0001$  \\
        $\beta_6$        &   0.0762  & 0.0230 &   3.3170  &   0.0009   &&   0.0807  & 0.0295 &   2.7330   &   0.0063 \\
		$\beta_7$        &   0.1659  & 0.0258 &   6.4400  & $<0.0001$  &&   0.1599  & 0.0284 &   5.6330   & $<0.0001$    \\\cmidrule{2-5}\cmidrule{7-10}
        {\it $\sigma$} \text{submodel}    &&&\\
        $\gamma_1$       &   0.6920 & 0.0368 & 18.7980 & $<0.0001$     &&   $-1.3634$ & 0.0160 & $-85.0720$ & $<0.0001$   \\
        $\gamma_2$       &   0.0069 & 0.0005 & 13.2020 & $<0.0001$     &&     0.0028  & 0.0002 &   13.5030  & $<0.0001$\\
        $\gamma_3$       &   0.1348 & 0.0257 & 5.2410  & $<0.0001$     &&     0.1554  & 0.0171 &   9.0650   & $<0.0001$\\
        $\gamma_4$       &   0.2521 & 0.0878 & 2.8710  &   0.0041      &&     0.2349  & 0.0720 &   3.2620   &   0.0011 \\
  \cmidrule{2-5}\cmidrule{7-10}\\
  {\it $\nu$} \text{submodel}    &&&\\
        $\lambda_1$      &  $-2.1870$ & 0.0225 & $-97.4200$ & $<0.0001$ && --   &  --  &   --   &  --     \\\cmidrule{2-5}\cmidrule{7-10}\\\hline
	\end{tabular}
\end{table}

The evidence above is further supported by the information criteria. The GBS model yields AIC and BIC values of 38697.64 and 38770.04, respectively, whereas the corresponding values for the BS model are 39114.62 and 39180.98. Since lower values of AIC and BIC indicate a better trade-off between model fit and complexity, both criteria favor the GBS model over the BS model for the rent data.

\begin{figure}[!htb]
	\centering
	\subfigure[Worm plot for GBS model]{\includegraphics[width=5cm,height=3.7cm]{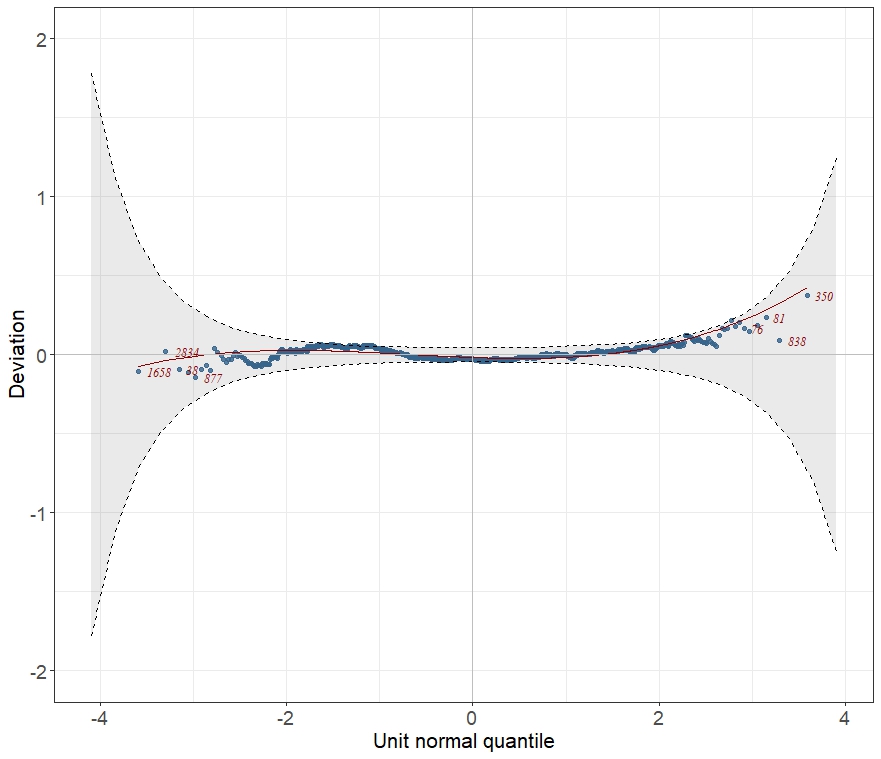}}
	\qquad
    \subfigure[Worm plot for BS model]
    {\includegraphics[width=5cm,height=3.7cm]{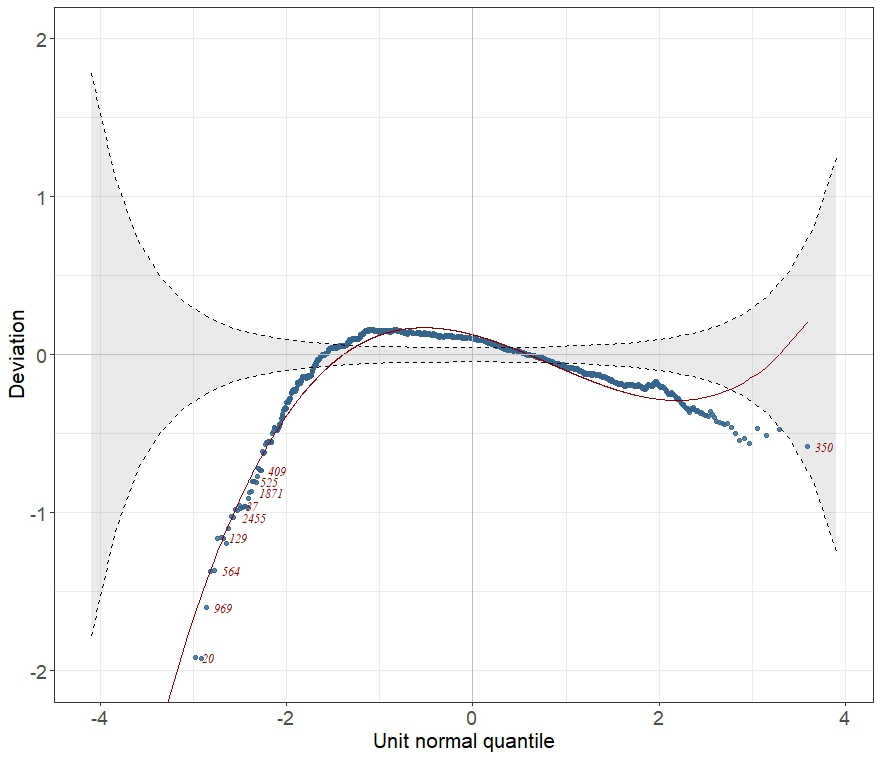}}\\
    \subfigure[Q-Q boxplot for GBS model]{\includegraphics[width=5cm,height=3.7cm]{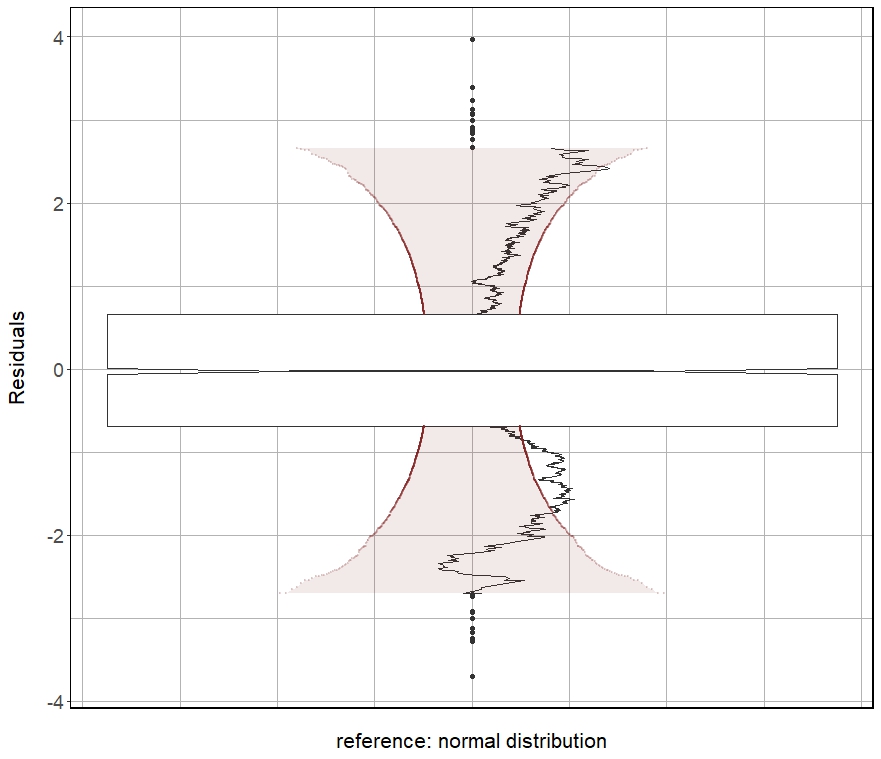}}
    \qquad
	\subfigure[Q-Q boxplot for BS model]{\includegraphics[width=5cm,height=3.7cm]{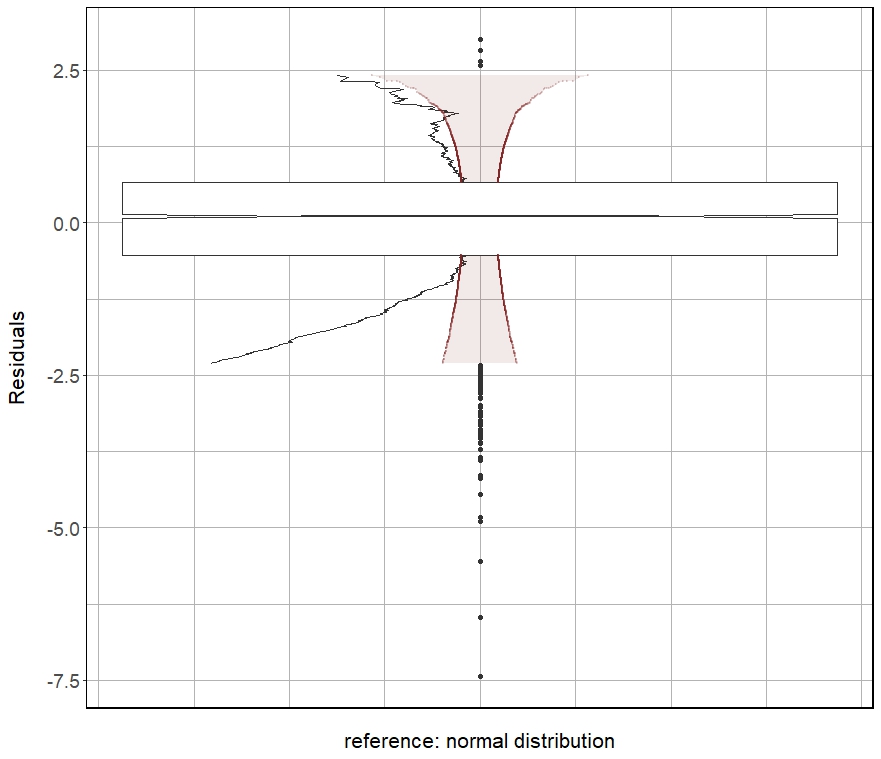}}\\
    \subfigure[Q-Q plot for GBS model]{\includegraphics[width=5cm,height=3.7cm]{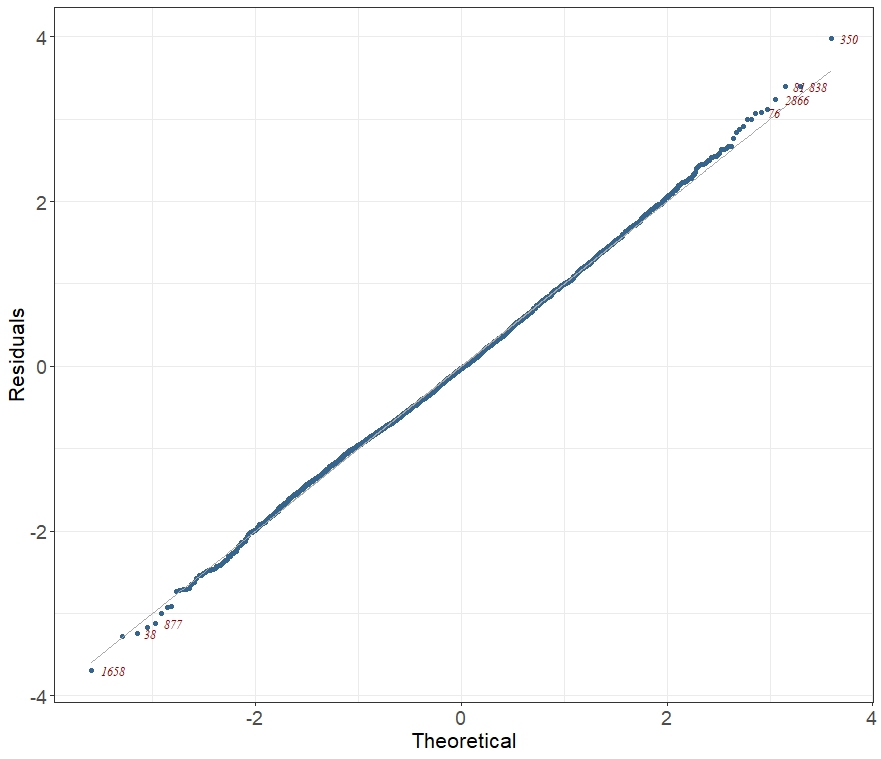}}
    \qquad
	\subfigure[Q-Q plot for BS model]{\includegraphics[width=5cm,height=3.7cm]{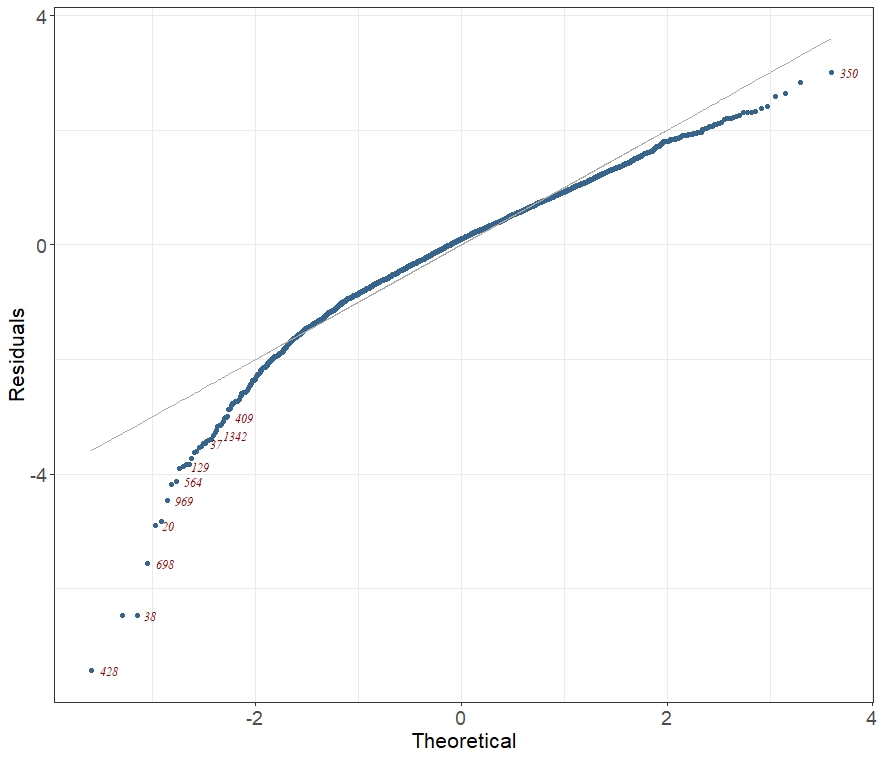}}\\
    \subfigure[Histogram of residuals with density estimator for GBS model]{\includegraphics[width=5cm,height=3.7cm]{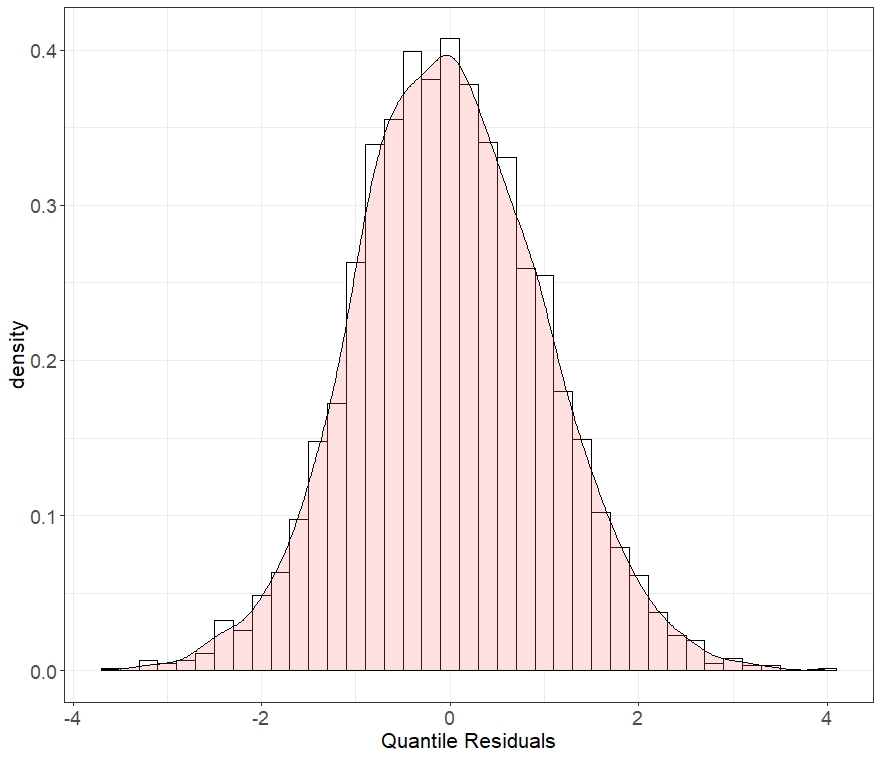}}
    \qquad
	\subfigure[Histogram of residuals with density estimator for BS model]{\includegraphics[width=5cm,height=3.7cm]{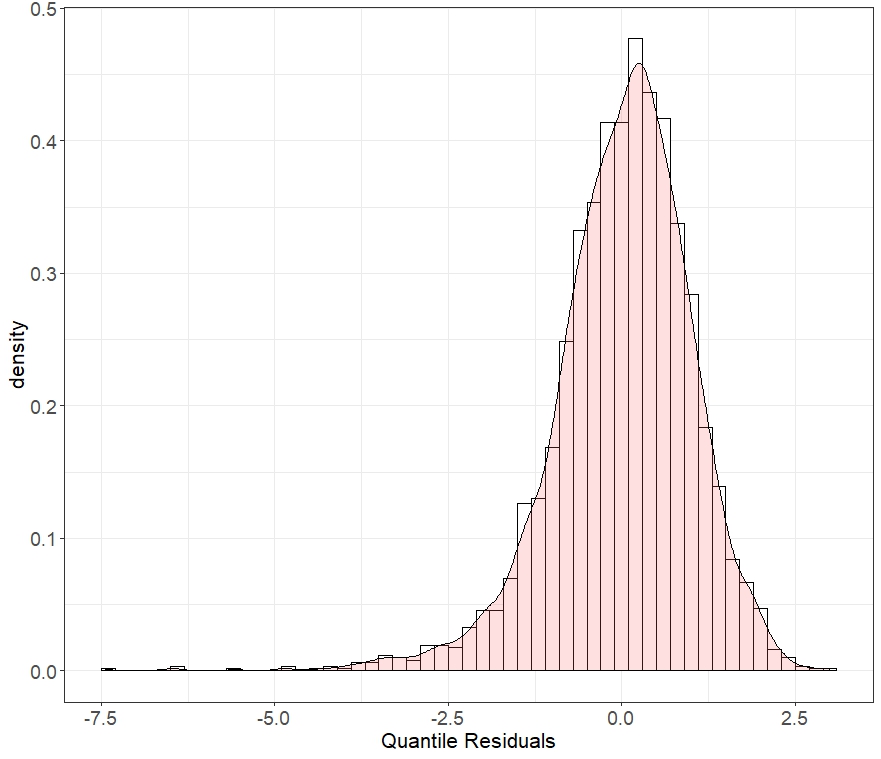}}
	\caption{Diagnostic plots based on the quantile residuals under the GBS and BS models -- rent data. } 
	\label{diagplots_2}
\end{figure}

Figure \ref{diagplots_2} presents diagnostic plots based on the quantile residuals for the fitted regression models. The worm plot for the BS model reveals a substantial lack of fit, with several points falling outside the upper and lower dotted curves and exhibiting clear patterns of systematic deviation. In contrast, the worm plot for the GBS model indicates a significantly better fit. Similar conclusions can be drawn from  the Q-Q boxplots, Q-Q plots, and histograms with density estimates of the quantile residuals. The Q-Q boxplot and histogram under the BS model show clear departures from the expected reference distribution, particularly a marked degree of asymmetry. The Q-Q plot of the residuals also reveals a lack of fit in the tails. On the other hand, these diagnostic plots under the GBS model indicate an adequate fit, with the quantile residuals displaying approximate symmetry and closely approximating the standard normal reference distribution. Hence, these diagnostic results suggest that the additional parameter $\nu$ in the GBS model provides considerable flexibility, allowing it to better  accommodate the distributional characteristics of the data. This finding highlights the important role of the parameter $\nu$ in improving the model's ability to capture deviations from the reference distribution and, consequently, achieve a more adequate fit.


Under the GBS regression model, with all other covariates kept constant, the effects of the covariates can be interpreted as follows.
An increase of 15 square meters in the living area is associated with a 17.94\% increase in the median monthly net rent, as $\exp(0.0110\times 15)=1.1794$. An increase of 20 years in the year of construction is associated with a 11.18\% increase in the median monthly net rent, since $\exp(0.0053 \times 20)=1.1118$. Furthermore, the median monthly net rent for flats in good locations is 9.12\% higher than that for flats in average locations, while the median monthly net rent for flats in top locations is 22.12\% higher than that for flats in average locations,  as indicated by $\exp(0.0873)=1.0912$ and $\exp(0.1998)=1.2212$, respectively. Flats with premium bathroom facilities have median monthly net rents that are 7.92\% higher than those with standard bathroom facilities, since $\exp(0.0762)=1.0792$. Finally, flats with premium kitchens have median monthly net rents that are 18.05\% higher than those with standard kitchens,  as $\exp(0.1659)=1.1805$.

\section{Concluding remarks}\label{sec7}

This paper proposes a new regression model based on a generalization of the BS distribution introduced by \cite{owen2006new}. The proposed model follows a specification similar to the GAMLSS approach. Hence, the models presented in this work allow covariate effects to be incorporated into all three distributional parameters, providing a flexible tool for modeling positive and asymmetric continuous data. The methodology introduced offers computational support for model fitting, inference, and diagnostic assessment through the package {\tt gamlss} for the software {\tt R}.

A particularly relevant feature of the proposed model is that one of its parameters corresponds directly to the median of the response variable. This parameterization offers a natural and interpretable approach to modeling covariate effects on the conditional median, which is especially advantageous for asymmetric data or when extreme observations are present, as the median is generally more robust to such values than the mean.

Parameter inference was conducted using maximum likelihood estimation. Hypothesis testing was performed using the Wald test, and diagnostic plots based on quantile residuals were evaluated, providing a comprehensive inferential framework for practical applications. Monte Carlo simulation results, which considered several scenarios, demonstrated that the maximum likelihood estimators exhibit desirable finite-sample properties, with bias decreasing and efficiency increasing as sample size grows. Additionally, the asymptotic standard error closely approximates the empirical standard error of the estimates. Furthermore, the empirical size of the Wald test was found to be close to the nominal significance level of 5\% for moderate to large sample sizes, supporting its use for inference.

The applications demonstrated the practical usefulness of the proposed models. In both cases, incorporating the additional parameter $\nu$ of the GBS distribution provided greater flexibility, resulting in an improved fit compared to the corresponding BS regression model. These results indicate that the additional parameter plays a crucial role in situations where the standard BS model fails to adequately describe the observed data.

As directions for future research, several extensions of the proposed framework can be considered. First, the behavior of maximum likelihood estimators in semiparametric and nonparametric GBS regression models could be investigated through simulation studies. Second, the development of diagnostic measures for identifying leverage and influential observations would be valuable for assessing model adequacy and detecting observations that may have a disproportionate impact on parameter estimates and inference. Third, alternative estimation procedures, including robust estimation methods, could be developed for GBS regression models to improve the model's robustness to outliers. 
Finally, the performance of alternative hypothesis testing procedures could be investigated, including the score and likelihood ratio tests, providing a broader inferential framework.

The code for all computational procedures in this article is available in an open-access repository at \url{https://github.com/braga0m/GBS_Regression_Models}.

	\paragraph*{Disclosure statement}
There are no conflicts of interest to disclose.

\paragraph*{Acknowledgements} This study was financed in part by the Coordenacao de Aperfeicoamento de Pessoal de Nivel Superior--Brazil (CAPES)--Finance Code 001 and the University of  Brasília (UnB).

\bibliographystyle{apalike}
\bibliography{ref_WTG}

\end{document}